\documentclass[reprint,amsmath,amssymb,aps,floatfix]{revtex4-2}

\usepackage{dcolumn}
\usepackage{bm}
\usepackage{amsmath}
\usepackage{diagbox}
\usepackage{appendix}
\usepackage{multirow}
\usepackage{hyperref}
\usepackage{acronym}   
\usepackage{subfigure}
\usepackage{graphicx}

\begin{document}
\preprint{APS/123-QED}

\title{Detection, sky localization and early warning for binary neutron star mergers by detectors located in China of different configurations in third generation detector network}
\author{YUFENG LI}
\affiliation{School of Physics and Technology, Wuhan University, Wuhan 430072, China}
\affiliation{SUPA, School of Physics and Astronomy, University of Glasgow, Glasgow G12 8QQ, United Kingdom}

\author{Ik Siong Heng}
\affiliation{SUPA, School of Physics and Astronomy, University of Glasgow, Glasgow G12 8QQ, United Kingdom}

\author{Man Leong Chan}
\email{mervync@phas.ubc.ca}
\affiliation{Department of Physics and Astronomy, The University of British Columbia, Vancouver, BC V6T 1Z4, Canada}

\author{XILONG FAN}
\email{xilong.fan@whu.edu.cn}
\affiliation{School of Physics and Technology, Wuhan University, Wuhan 430072, China}

\author{Lijun Gou}
\affiliation{Key Laboratory for Computational Astrophysics, National Astronomical Observatories, Chinese Academy of Sciences, Beijing 100101, China}
\affiliation{School of Astronomy and Space Science, University of Chinese Academy of Sciences, Beijing 100049, China}
\begin{abstract}
This work shows the results of an evaluation of the impact that a detector located in China, with a noise budget comparable to that of a proposed high-frequency detector with a 20 km arm length, an \ac{ET} or a \ac{CE}, could have on the network of ET-CE in terms of detection rate, localization, and providing early warning alert for simulated \ac{BNS}s. The results indicate that a three-detector network including a Chinese detector could identify at least 4.4\% more \ac{BNS} mergers than an ET-CE network alone. The localization uncertainty could be reduced by a factor of more than 5 on average compared to the ET-CE network. With a three-detector network involving a Chinese detector, up to 89\% of \ac{BNS} mergers could be located within 10 square degrees of the sky 10 minutes prior to the merger. The assessment suggests that the potential for early warning signals is highest when the Chinese detector is similar to ET, whereas the sources are detected with the highest signal-to-noise ratio and localized to the smallest regions when the detector is more akin to CE. Interestingly, the C20N network (comprising ET+CE+C20) can achieve comparable localization performance as the ET network while outperforming the ETCN network (featuring the ET+CE+ an ET-like detector in China) in terms of detection capabilities, especially at large distances, indicating that adding a 20 km kilohertz detector in China to ET-CE network would make significant contributions at least as adding an ET-like detector in China to multi-messenger astronomy for almost all BNS observations.
\end{abstract}
\maketitle
\acrodef{GW}[GW]{gravitational wave}
\acrodef{EM}[EM]{electromagnetic}
\acrodef{BNS}[BNS]{binary neutron star}
\acrodef{NSBH}[NSBH]{neutron star - black hole}
\acrodef{SNR}[SNR]{signal to noise ratio}
\acrodef{ET}[ET]{Einstein Telescope}
\acrodef{CE}[CE]{Cosmic Explorer}
\acrodef{ASD}[ASD]{amplitude spectral density}
\section{Introduction}\label{intro}
 Two years after the first detection of \ac{GW} on September 14, 2015 [\cite{Abbott2016b, Abbott2016c, Abbott2016a}], a \ac{GW} signal from a binary neutron star inspiral was observed by Advanced LIGO and Advanced Virgo on August 17, 2017, leading to the first-ever joint detection of gravitational and electromagnetic radiation from a single source~[\onlinecite{Abbott2017}]. 
 The identification of the correlated gamma-ray burst by ~[\onlinecite{Fermi2017}], along with the subsequent \ac{EM} follow-up campaign, heralded the era of multi-messenger astronomy, synergized with \acp{GW} [\onlinecite{Abbott2017}]. The case of \ac{GW}170817~[\onlinecite{Abbott2017}] demonstrated the tremendous scientific potential of successfully observing the associated \ac{EM} counterparts of a \ac{GW} event. Observing an \ac{EM} counterpart offers several advantages. Establishing a link between the \ac{GW} trigger and its progenitor can yield valuable insights into the progenitor and its surrounding environment~[\onlinecite{Mottin2011, Bloom2009, Kanner2008}]. Moreover, independently measuring the redshift of the source emitting the \ac{GW} signal via the observation of an \ac{EM} counterpart facilitates cosmological examinations and constraining the equation of state of dark energy~[\onlinecite{Schutz1986, Sathyaprakash2010, Abbott2017d}]. Hence, comprehending how to optimize the combined efficiency of \ac{GW}s and their corresponding electromagnetic counterparts is crucial for maximizing scientific advancements.\\
\indent Localization uncertainty and early warning capability (the ability to detect signals prior to mergers with reasonable localization uncertainty ~[\onlinecite{Cannon2012}]) are pivotal factors directly influencing the effectiveness of combined gravitational wave (GW) and electromagnetic (EM) observations. Second-generation detectors like Advanced LIGO~[\onlinecite{LIGO2015}], operating at their design sensitivity, are projected to detect only $10^{-5}$ of all binary neutron star (BNS) mergers. Moreover, the median sky localization area (90\% credible region) for compact binary coalescences during the third observation run of the LIGO and Virgo collaborations was a few hundred square degrees ~[\onlinecite{Baibhav2019}][\onlinecite{Abbott2020}]. Therefore, the sensitivity of current-generation detectors is constrained in their capacity to conduct a substantial number of GW-EM joint observations. However, with the emergence of third-generation detectors such as the Einstein Telescope (ET)~[\onlinecite{Punturo2010}], a proposed underground detector in Europe (more details available on the ET website(http://www.et-gw.eu), and the Cosmic Explorer (CE)~[\onlinecite{Reitze2019}], a planned next-generation GW observatory in the United States(https://cosmicexplorer.org), BNS detection rates could improve by up to 3-4 orders of magnitude~[\onlinecite{Branchesi2023}, \onlinecite{Gupta2023}]. These detectors are expected to detect about 50\% of all BNS mergers and localize the majority of them to $\leq 200 \mathrm{Mpc}$ with uncertainties of $\mathcal{O}(1) \operatorname{deg}^{2}$, and $\leq 100 \mathrm{deg}^{2}$ for most BNS mergers $\leq 1600 \mathrm{Mpc}$, as reported by~[\onlinecite{Chan2018}]. By incorporating a third detector into the network comprising \ac{ET}-\ac{CE}, the angular resolution of gravitational wave events will experience a significant enhancement. This enhancement stems from the fact that the localization uncertainty is inversely proportional to the area enclosed by the three detectors~[\onlinecite{2010PhRvD..81h2001W}]. Consequently, a larger area encompassed by the three detectors is favored to minimize localization errors for the three-detector network. \\
\indent Recent research has investigated the potential of incorporating a detector in Australia into the \ac{ET}-\ac{CE} network, as proposed in studies by [\onlinecite{Barriga2010, 2011CQGra..28j5021F, Hu2015, Rudenko2020}]. In a similar vein, as detailed in our previous work~[\onlinecite{2021arXiv210907389L}], we investigated the detection capabilities, localization uncertainty, and early warning performance of a three-detector network (comprising \ac{ET}, \ac{CE}, and a comparable detector in Australia) during the third-generation detector era for \ac{BNS} mergers. We discovered that the extensive baseline between these detectors enabled the localization of the majority of sources within 0.25 deg$^2$ up to distances of 200 Mpc. Moreover, the prolonged in-band duration enabled by third-generation detectors could facilitate the accumulation of sufficient signal-to-noise ratio for a potential \ac{GW} signal to achieve statistical significance before the merger occurs. Consequently, this could improve the likelihood of early warning and subsequent \ac{EM} follow-up observations.\\
\indent Based on the preceding discussion, it becomes evident that establishing a global network of gravitational wave detectors, especially in diverse geographic locations, is a crucial step towards enhancing the precision of gravitational wave source localization. Given our interest in China's contributions to the field of gravitational wave localization, it's worth noting that its geographic location could create a vast triangular area with Europe and the United States. This positioning holds the potential to significantly augment both the detection rate and localization accuracy when compared to the ET-CE network. \\
\indent Therefore, assessing the performance of gravitational wave detectors in China serves not only the demands of scientific inquiry but also the advancement of technological innovation and high-tech sectors within China. Additionally, it fosters collaboration and growth in global scientific research endeavors. Motivated by these considerations, our objective is to examine the detection capabilities, localization accuracy, and early warning efficacy of a three-detector network comprising third-generation \ac{GW} detectors—namely \ac{ET}, \ac{CE}, and a comparable detector in China—for \ac{BNS} mergers. This assessment aims to gauge the contribution of Chinese detectors to the global \ac{GW} detector network in advance. Such insight is crucial for determining whether China should establish a ground-based \ac{GW} detector in the future and what configuration would be most advantageous for localization efforts.\\
\indent Prior studies conducted by [\onlinecite{Hu2015}] and [\onlinecite{Rudenko2020}] have investigated the influence of a hypothetical detector in China on the network's capacity to localize gravitational wave sources. Utilizing three Figures of Merit—reconstruction polarization capability, localization accuracy, and parameter accuracy—they concluded that the southern region of China would offer improved performance when integrated into a network comprising advanced LIGO Hanford, advanced LIGO Livingston, advanced Virgo, and KAGRA. We have chosen Wuhan (30.52$^{\circ}$N, -245.69$^{\circ}$E) in Hubei province as the testing ground for our virtual \ac{GW} detectors. Its central position in China, far from coastal areas and seismic zones, along with its flat terrain and stable underground structures, render it an optimal site for ground-based detector. These attributes reduce the influence of earthquakes and environmental noise on the detectors, thereby enhancing their sensitivity and therefore localization accuracy. Moreover, the Wuhan city government is capable of providing ample land for constructing a \ac{GW} detector. Furthermore, Wuhan's suitability has previously been evaluated for its potential to constrain string cosmology in a study by [\onlinecite{Li2019}]. Our assessment of its localization performance will provide additional insights into the feasibility of Wuhan as a prospective site for a future ground-based gravitational wave detector in China.\\
\indent In China, there have been ongoing discussions regarding the configuration of potential future ground-based detectors, with one particularly unique proposal being the kilohertz detector. Distinguished by its exceptional sensitivity at kilohertz frequencies, this detector is poised to bridge the gap for future gravitational wave detection within this frequency range. Notably, it holds the promise of observing neutron star post-merger oscillations—an occurrence arising from the energy released during neutron star mergers, which induces high-frequency oscillations in the resulting remnant~[\onlinecite{Martynov2019}]. Kilohertz detectors possess the capability to detect and analyze these oscillations, thereby furnishing invaluable insights into the properties and structure of neutron stars, the equation of state of nuclear matter, and the remnant's role as the central engine for energetic electromagnetic emissions. [\onlinecite{2019PhRvD..99j2004M}] suggested that an optimal arm length of approximately 20 km is necessary for observing neutron star post-merger oscillations. To assess the viability of establishing such a 20 km kilohertz detector in China in the future, various aspects of the detector's performance need consideration, including sensitivity, economic and technical feasibility, signal-to-noise ratio, and localization accuracy. In our previous work~[\onlinecite{Li2019}], we evaluated the constraining capability of a 20 km detector in Wuhan for string cosmology. Furthermore, assessing the localization accuracy of a 20 km detector in Wuhan will further inform the discussion on whether China should proceed with establishing such a kilohertz detector.\\
\begin{table*}
  \centering
  \begin{tabular}{cccc}
      \hline
      \hline
        &   China & LIGO Hanford & Italy \\
        &  ( $-245.69^{\circ}$E,$30.52^{\circ}$N)&  ($-119.41^{\circ}$E,  $46.45^{\circ}$N)&($10.4^{\circ}$E,  $43.7^{\circ}$N) \\
        \hline
       ETCN& ET-C & CE & ET  \\  
       \hline
       CECN& CE-C & CE & ET \\  
       \hline
       C20N & C20 & CE & ET \\  
      \hline
       \hline
  \end{tabular}
  \caption{Tested networks in this paper, the first column represents the abbreviation for network, where N represents a network. Three kinds of detectors located in China, respectively an ET-like detector~(ET-C), a CE-like detector~(CE-C), and a high-frequency detector~(C20) is assumed to operate with ET and CE to form a three-detector network. }\label{GWD}
  \end{table*} 
\indent By simulating a 20 km kilohertz detector (hereafter referred to as C20) in Wuhan, forming part of the C20N network (comprising ET+CE+C20, as detailed in Table~\ref{GWD}), we conducted tests to evaluate the network's performance and compared it with the ET-CE network. Our study aims to uncover the added value of including C20 in terms of localization performance compared to the ET-CE network. Furthermore, to delve deeper into the influence of various configurations on localization capability in Wuhan, we performed simulations of two additional detectors—an ET-like detector (ET-C) and a CE-like detector (CE-C). Consequently, two testing networks (ETCN: ET-C+ET+CE; CECN: CE-C+ET+CE) were established by combining these simulated detectors with ET and CE, as outlined in Table~\ref{GWD}.\\
\indent Moreover, we utilized the same sources as in [\onlinecite{2021arXiv210907389L}], facilitating a direct quantitative comparison of detection rates, localization accuracy, and early warning performance between deploying the third detector in Australia and China. These sources comprise \ac{BNS} mergers at fixed distances (100 sources each at 40 Mpc, 200 Mpc, 400 Mpc, 800 Mpc, and 1600 Mpc) and \ac{BNS} mergers following an assumed astrophysical population (500 sources with a delay time distribution) up to a redshift of $\leq$ 2. The information regarding our simulated sources is outlined in Table~\ref{source_info_table}, while the redshift distribution of \ac{BNS} mergers, following a delay time distribution, is depicted in Figure~\ref{z_distribution}. We adopt the same methodology as employed in [\onlinecite{2021arXiv210907389L}], which involves calculating \ac{SNR} and localization uncertainty using the Fisher Matrix. In the interest of brevity, our focus remains on presenting the results and comparing them with previous findings. For more comprehensive insights into the methods and simulated sources, please refer to [\onlinecite{2021arXiv210907389L}].\\
\indent This paper is organized as follows. Section~\ref{detector} presents the basic information of third-generation ground-based detectors. Section~\ref{results} presents the results of our analysis. In Section~\ref{discussion}, we discuss our findings and provide concluding remarks.

\section{Third Generation Detector}\label{detector}
\begin{figure}
\centerline{\includegraphics[scale = 0.2]{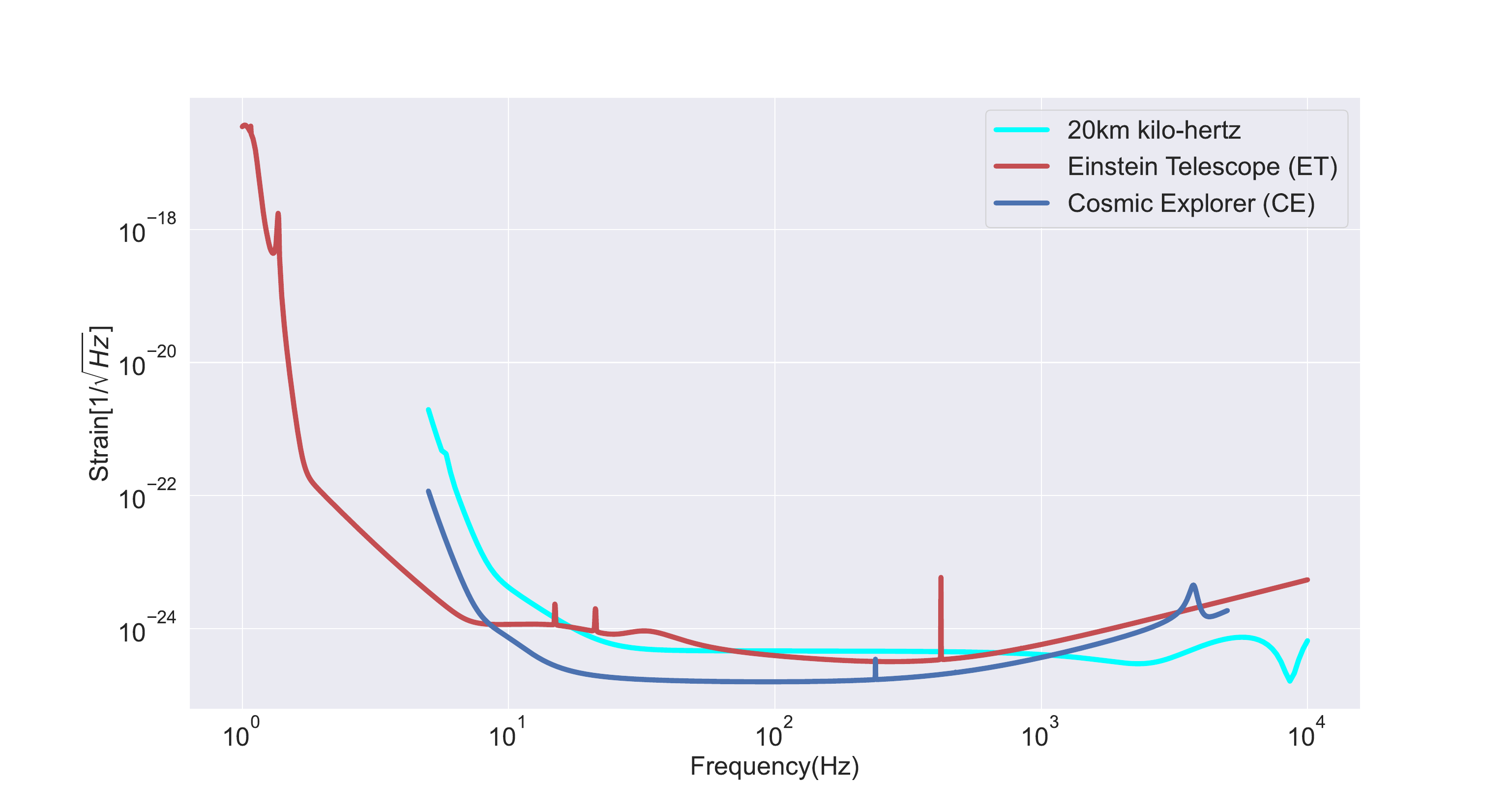}}
\caption{The amplitude spectral density against frequency for the Einstein Telescope (ET), Cosmic Explorer(CE), and 20\,km kilohertz detector.}\label{strain}
\end{figure}

As of the conclusion of their third observing run, Advanced LIGO and Advanced Virgo collectively detected 90 compact binary coalescences, as documented in~[\onlinecite{2021arXiv211103606T}]. These detections carry significant implications for our comprehension of compact binary coalescences and cosmology. Nevertheless, current ground-based detectors are not without limitations, prompting proposals to enhance sensitivity and discover more sources with third-generation ground-based \ac{GW} detectors projected for the 2030s. The two primary designs for these third-generation \ac{GW} observatories are \ac{CE} and \ac{ET}. The \ac{CE} design mirrors the current second-generation \ac{GW} detectors, featuring an L-shaped configuration housing a single interferometer with extended arm lengths of up to 40 km, as delineated in~[\onlinecite{2017CQGra..34d4001A}] and~[\onlinecite{2015PhRvD..91h2001D}]. Conversely, the proposed \ac{ET} comprises an underground infrastructure incorporating three interferometers with arm lengths of 10 km, arranged in an equilateral triangular configuration, as elaborated in~[\onlinecite{2010CQGra..27a5003H}].\\
\indent Figure~\ref{strain} depicts the sensitivity curves, expressed as the \ac{ASD} as a function of frequency, for \ac{ET}, \ac{CE}, and C20. Notably, \ac{ET} showcases superior sensitivity in the low-frequency band ($<\sim$9 Hz) compared to \ac{CE}. This enhancement primarily stems from the substantial reduction of thermal noise achieved by operating the mirrors at cryogenic temperatures as low as 10K~[\onlinecite{2010CQGra..27s4002P, 2011ET-0106C-10}]. The heightened low-frequency sensitivity of \ac{ET} enables earlier detection of inspiralling \ac{GW} sources and the potential for more early warning alerts. With its significantly longer arm length, \ac{CE} offers markedly better sensitivity than the other two detectors across nearly the entire frequency band (10 Hz - 1000 Hz), thereby greatly enhancing its ability to explore the depths of the universe and enabling it to detect more sources compared to \ac{ET}. On the other hand, C20 exhibits excellent sensitivity in the high-frequency band (>1000 Hz) and comparable sensitivity to \ac{ET} in the frequency range of approximately 15 Hz to 1000 Hz. Beyond 1 kHz, the sensitivity of \ac{GW} detectors is typically influenced by the combination of quantum shot noise and classical noises, such as coating thermal noise and gas phase noise. To address these constraints and bolster sensitivity beyond 1 kHz, [\onlinecite{2019PhRvD..99j2004M}] recommended augmenting input power and introducing squeezed states of light, thereby proposing the concept of a 20 km high-frequency detector. The heightened sensitivity of the 20 km detector in this frequency range proves advantageous for detecting post-merger phases of \ac{BNS} mergers and holds potential for achieving commendable localization performance, particularly at the time of the merger. Nonetheless, the sensitivity of the 20 km detector experiences a pronounced decline below 15 Hz, limiting its capacity to detect inspiralling \ac{GW} sources at earlier stages, a capability where \ac{ET} excels.\\
\indent The sensitivity of a gravitational wave (GW) detector to an astrophysical source hinges on two key factors: its sensitivity across frequencies, quantified by its amplitude spectral density, and its antenna pattern, which is contingent upon the detector's geographical position, arm configuration, and the angle between its arms. References such as [\onlinecite{Regimbau2012}] and [\onlinecite{2021arXiv210907389L}] provide details on the antenna patterns for both the L-shaped detectors (CE, CE-C, C20) and the equilateral triangular-shaped detectors (ET, ET-C). Moreover, for the three virtual detectors situated in Wuhan (ET-C, CE-C, and C20), we adopted the same arm orientation as KAGRA, as done previously in [\onlinecite{Li2019}]. This choice is supported by studies such as [\onlinecite{2021arXiv210812698R}], which indicate that the localization criterion is not significantly impacted by the orientation angle of the detector.
\section{Results}\label{results}
\subsection{Simulation I: BNS mergers following delay time distribution}
\subsubsection{Detection rates}
\begin{figure}
\centering
\includegraphics[scale  = 0.185]{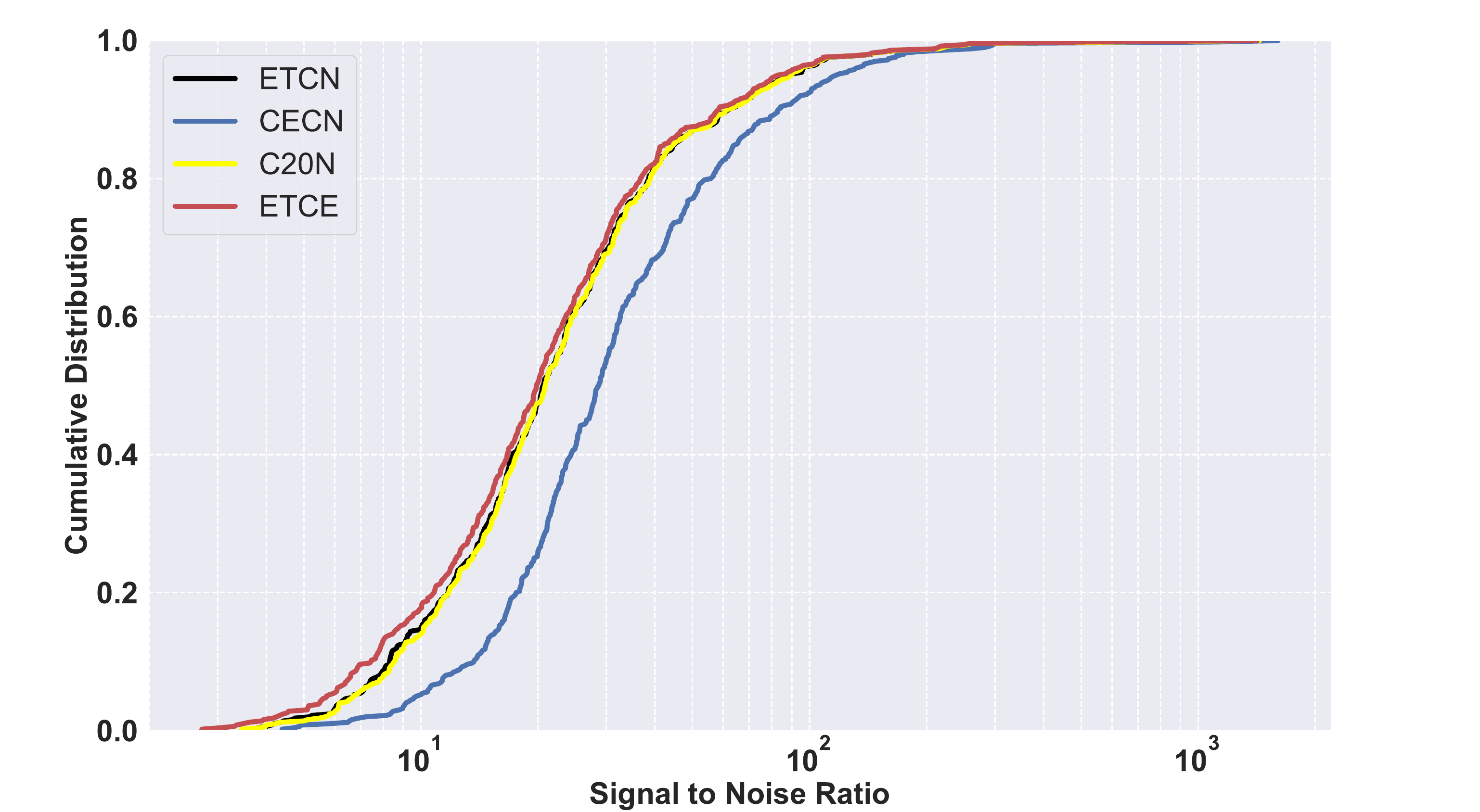}
\caption{The cumulative distribution of \ac{SNR} of tested three detector network configurations and also ET-CE network for the \ac{BNS}s in Simulation I.}\label{snr_hist}
\end{figure} 
\indent The optimal signal-to-noise ratio (\ac{SNR})~ (details see Eq(11) of [\onlinecite{2021arXiv210907389L}]) for 500 simulated sources, as observed by gravitational wave (\ac{GW}) detector networks listed in Table~\ref{GWD}, has been calculated. Assuming a network \ac{SNR} of no less than 8 for a source to be detectable, the analysis reveals detection rates of 91.4\%, 98\%, 92.2\%, and 87\% for sources by the ETCN, CECN, C20N, and \ac{ET}-\ac{CE} networks, respectively, as summarized in the second row of Table~\ref{loct_3G}. Furthermore, the table presents results for the ET-CE network combined with either a CE-like detector in Australia (1ET2CE) or an ET-like detector in Australia (2ET1CE), showcased in the last two columns. These findings are derived using the same sources as those examined in [\onlinecite{2021arXiv210907389L}], facilitating a direct comparison of the outcomes.\\
\indent We observed that the inclusion of a detector in China resulted in a minimum of 4.4\% more detectable sources compared to the ET-CE network. Among the three detector networks incorporating a Chinese detector, the CECN network exhibited superior performance in source detection compared to the other two networks, while the C20N network, overall, detected 0.8\% more sources than the ETCN network. Upon comparing with 1ET2CE and 2ET1CE configurations, we found that the ETCN network detected 1.6\% more sources than the 2ET1CE network, and the CECN network detected 1\% more sources than the 1ET2CE network. These results suggest that augmenting the ET-CE network with an ET-like (or CE-like) detector in China could yield a higher detection rate compared to integrating a similar detector in Australia. This difference is primarily attributed to variations in antenna response. There are two key aspects to this: Firstly, the geographic positions of the detectors determine the area of the triangle formed by the network and the baseline lengths. A detector in China alters the network geometry, resulting in a larger detector triangle area and different baseline lengths, which enhances the network's triangulation and detection capabilities. Secondly, the orientation of the detector arms affects their sensitivity to gravitational waves from different sources. A detector in China would have different arm orientations compared to one in Australia, leading to variations in the antenna response and, consequently, differences in sensitivity to sources from various parts of the sky. Therefore, the overall improvement in detection rates is largely due to the differing antenna responses influenced by both the geometric configuration of the network and the orientation of the detector arms.\\
\indent Figure~\ref{snr_hist} illustrates the cumulative distribution of signal-to-noise ratio (\ac{SNR}) for simulated sources detected by networks equipped with detectors in China. The \ac{SNR} distribution of the CECN network tends to skew towards higher values, contrasting with the ETCN network and C20N network, whose \ac{SNR} distributions are nearly identical. However, in quantitative terms, the C20N network outperformed the ETCN network by detecting 0.8\% more sources overall.\\
\begin{figure}
\centerline{\includegraphics[scale = 0.185]{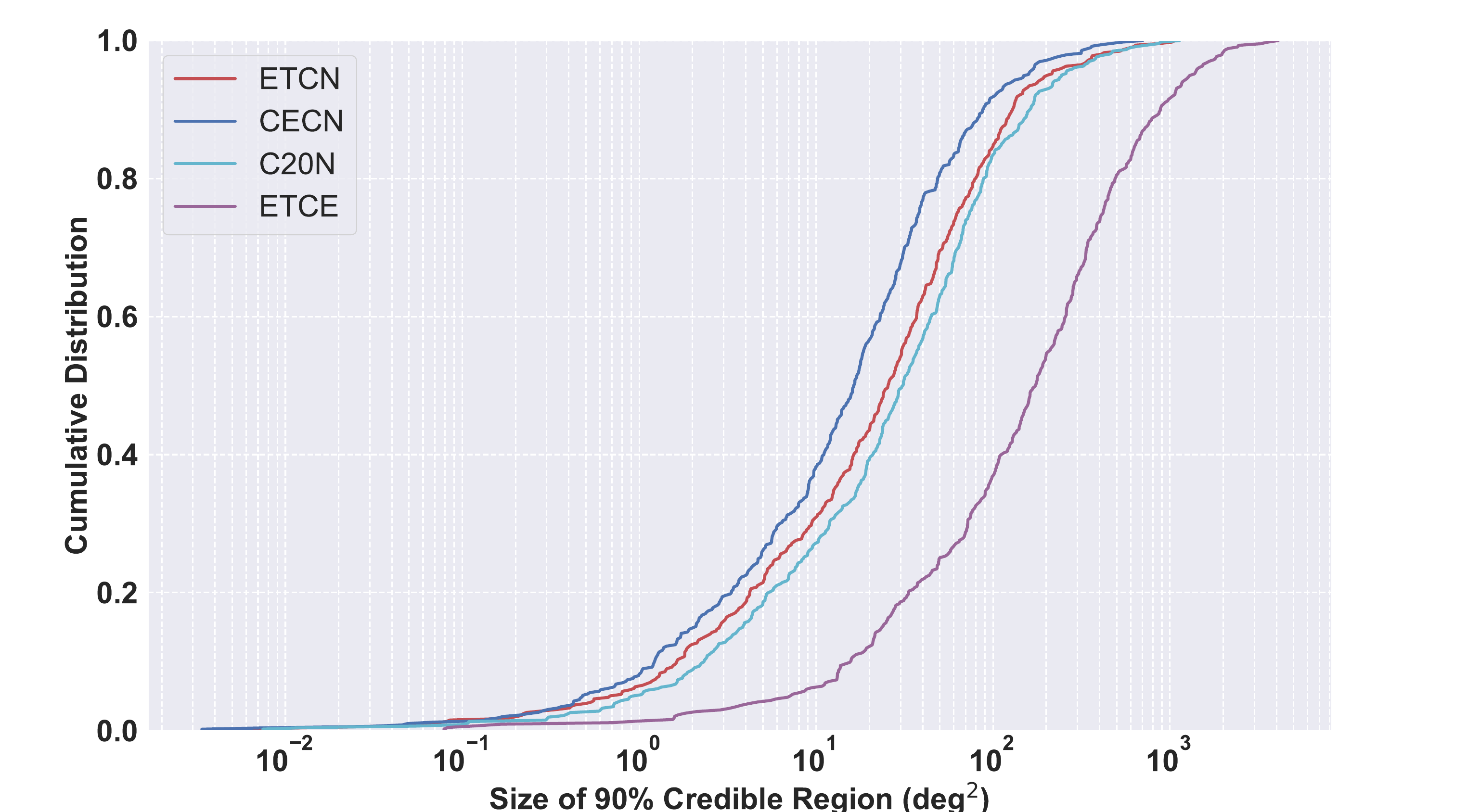}}
\caption{Cumulative distribution of localization uncertainty for the \ac{BNS}s in  Simulation I. The vertical axis shows the cumulative fraction.The horizontal axis is the size of 90\% credible region in deg$^2$.}\label{er}
\end{figure}
\begin{table}
  \centering
  \begin{tabular}{ccccccc}
      \hline
      \hline
        & ETCN & CECN & C20N & ETCE &1ET2CE & 2ET1CE \\
      \hline  
        & 91.4\% & 98\% & 92.2\% & 87\% & 97\% & 89.8\%\\            
       \hline
        90\% & $128.86$ & $87.43$ & $161.51$ & $881.88$ & $56.77$ &  $91.79$\\
        50\%  & $25.49$& $16.29$ & $30.82$    & $173.41$& $12.54$ &  $17.57$\\
        10\%  & $1.63$ & $1.22$ & $2.33$      & $15.62$&  $1.17$  &  $1.37$\\     
      \hline
      \hline
  \end{tabular}
  \caption{Values show the 90\% credible regions of localization uncertainty in  deg$^2$ for detectable simulated the \ac{BNS}s in  Simulation I. The second row denotes the fraction of detectable sources for each detector network with SNR$\geq 8$.  90\%, 50\% and 10\% in the first column respectively represents the best localized 90\%, 50\% and 10\% of the detectable sources. The fifth column shows the results for ET-CE network, the sixth column shows the results for ET-CE network plus a CE-like detector in Australia, the last column shows the results for ET-CE network plus an ET-like detector in Australia. }\label{loct_3G}
  \end{table} 
\subsubsection{Localization performance}
\indent  Figure~\ref{er} illustrates the cumulative distribution of localization uncertainties across various networks. All localization uncertainties in this paper refer to the 90\% credible interval of estimated localization uncertainty. Table~\ref{loct_3G} further details the localization uncertainty for the top 90\%, 50\%, and 10\% most accurately localized detectable sources, comparing networks with and without a detector in China alongside the \ac{ET}-\ac{CE} network for reference. Key observations include: (1) The \ac{ET}-\ac{CE} network achieves localization of the top 90\% sources within approximately 882 deg$^2$. With the addition of a Chinese detector, this uncertainty shrinks dramatically to $\le$ 162 deg$^2$, representing less than one-fifth of the original sky error. (2) In horizontal comparison with networks incorporating an Australian detector, the ETCN network exhibits localization uncertainty approximately 1.4 times greater than the 2ET1CE network for the top 90\% of sources. Similarly, the CECN network's uncertainty is approximately 1.54 times greater than that of the 1ET2CE network. This implies that augmenting the \ac{ET}-\ac{CE} network with an Australian detector leads to more precise localization for identical \ac{BNS} mergers compared to integrating a detector in China~(It should be noted that in the case of China, the number of sources that meet the detection criteria is relatively higher, which means that some sources with larger localization errors are also included in the statistics. In contrast, for Australia, the sources that are included in the statistics mostly have smaller localization errors. Thus the Chinese networks give better detectability but worse localization than the Australian
network.).
(3) Among networks featuring a Chinese detector, the CECN network demonstrates the lowest overall localization uncertainty. Specifically, the localization uncertainty of the ETCN (C20N) network is roughly 1.5 (2) times larger than that of the CECN network.
\subsubsection{Early Warning}
\begin{figure}[h]
\subfigure[ETCN network] { \label{fig:a} 
\includegraphics[width=1\columnwidth]{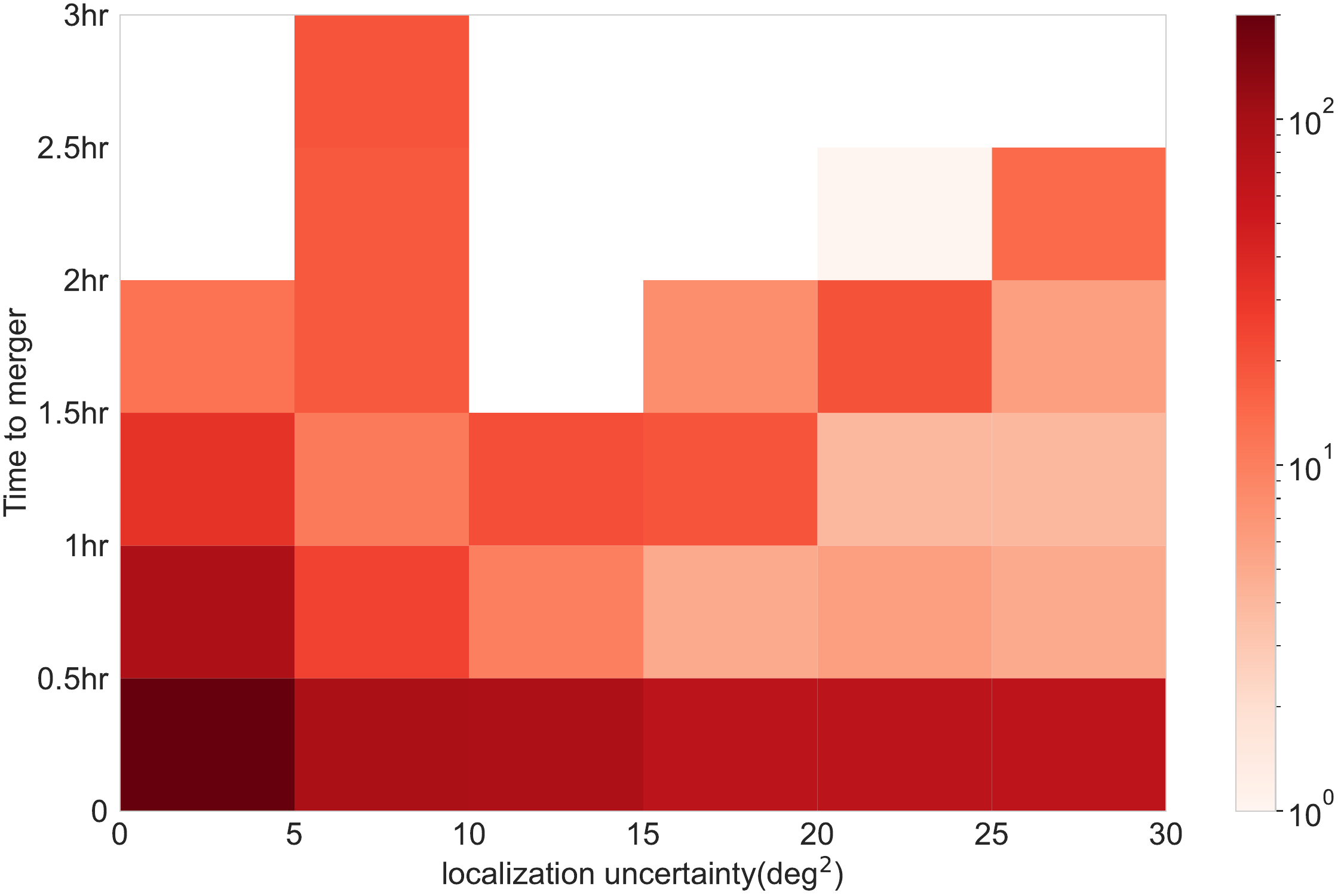}} 
\subfigure[CECN network] { \label{fig:b} 
\includegraphics[width=1\columnwidth]{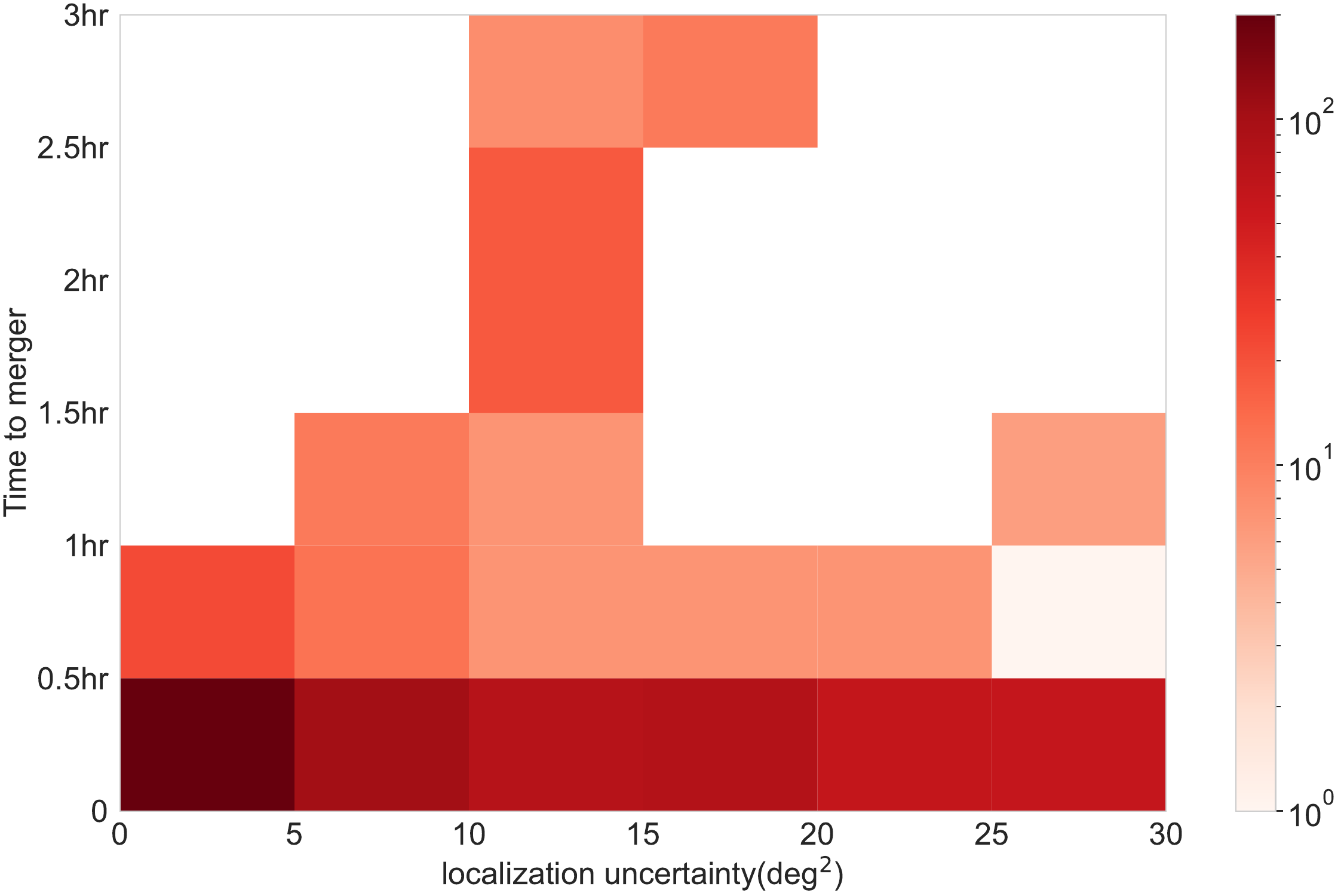}} 
\subfigure[C20N network] { \label{fig:c} 
\includegraphics[width=1\columnwidth]{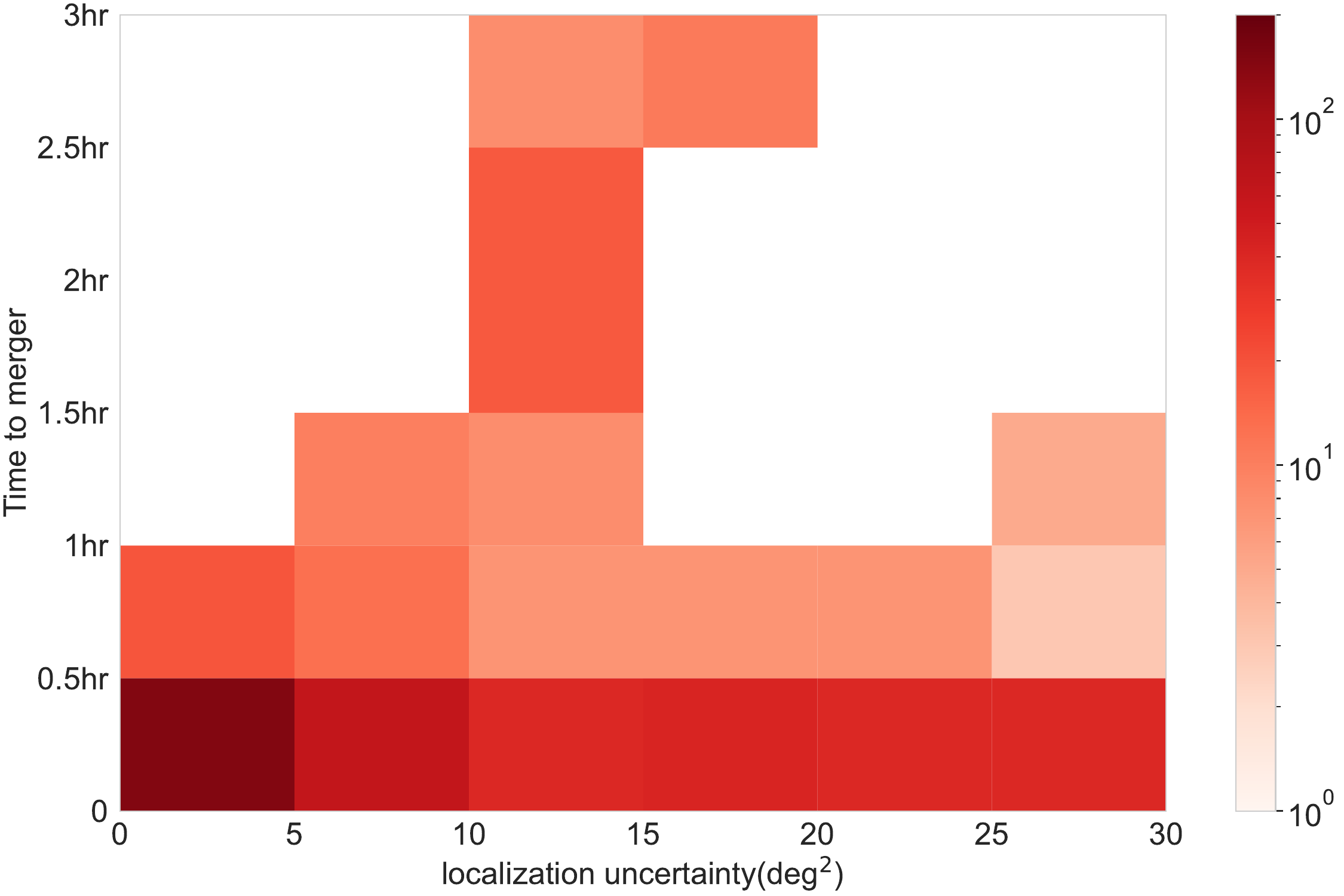}} 
\caption{Two-dimensional histograms showing the distributions of the time to merger for the \ac{BNS}s in Simulation I. The horizontal axis of every subplot is the size of the 90\% credible region, and the vertical axis shows the time to merger. The color represents the number of sources that achieve the early warning criteria with the localization requirement being $\leq$ 30 deg$^2$.}\label{EW_3G}
\end{figure}

\indent If the inspiral signal from a binary neutron star (BNS) merger can attain a sufficiently robust signal-to-noise ratio (SNR) and be localized within a reasonable portion of the sky prior to the actual merger, it opens the possibility of issuing an early warning alert to facilitate observations of any associated electromagnetic counterparts. In our investigation, we operate under the assumption that for an early warning to be issued, the source must satisfy two criteria: firstly, the network SNR for the source must reach a minimum of 8, and secondly, the localization uncertainty should not exceed a predefined sky area (either 30 or 100 deg$^2$).\\
\indent The two-dimensional histogram depicting the distribution of time to merger versus the size of the 90\% credible region (less than 30 deg$^2$) when signals have accumulated a network \ac{SNR} of $\geq 8$ for the binary neutron stars (\ac{BNS}s) in Simulation I is presented in Figure~\ref{EW_3G}. Among the detector networks, ETCN, CECN, and C20N respectively detect 54\%, 67\%, and 49\% of sources with an SNR of $\geq$ 8, successfully localizing them within 30 deg$^2$ at the time of merger. Notably, the time to merger distribution between CECN and C20N networks exhibits remarkable similarity, with both networks capable of detecting the majority of sources within 3 hours before the merger. Furthermore, ETCN network, overall, demonstrates an earlier detection capability compared to the other two networks. The results pertaining to the maximum allowable region of 100 deg$^2$ are provided in the appendix.

\subsection{Simulation II: BNS mergers at fixed distances}
\subsubsection{Detection rates}
Similarly, with a network \ac{SNR} threshold set at 8 for detection, both the CECN and C20N networks demonstrated the capability to detect all simulated sources across all tested distances, ranging from 40 Mpc to 1600 Mpc. However, for the ETCN network, three signals at 1600 Mpc remained undetected, while the \ac{ET}-\ac{CE} network missed five signals at the same distance.
\subsubsection{Localization performance}
\begin{table*}
\centering
  \begin{tabular}{p{1.5cm}p{1cm}cccccc}
      \hline
      \hline
       &  & ETCN & CECN & C20N & ETCE & 1ET2CE & 2ET1CE\\
       \hline
          & $30\mathrm{deg^2}$ & 100\% & 100\% & 100\%  & 100\% & 100\% & 100\%\\
    200Mpc& $10\mathrm{deg^2}$ & 100\% & 100\% & 100\%  & 99\%  & 100\% & 100\%\\
          & $5\mathrm{deg^2}$ & 100\% & 100\% & 100\%   & 98\%  & 100\% & 100\%\\
          & $1\mathrm{deg^2}$ & 98\% & 97\% & 96\%      & 72\%  & 100\% & 100\%\\
       \hline
          & $30\mathrm{deg^2}$ & 100\% & 100\% & 100\%  & 99\%  & 100\% & 100\%\\
    400Mpc& $10\mathrm{deg^2}$ & 100\% & 100\% & 100\%  & 92\%  & 100\% & 100\%\\
          & $5\mathrm{deg^2}$ & 99\%& 100\% & 98\%      & 76\%  & 100\% & 100\%\\
          & $1\mathrm{deg^2}$ &74\%& 92\% & 69\%        & 10\%  & 95\%  & 89\%\\
       \hline
          & $30\mathrm{deg^2}$ & 100\% & 100\% & 100\%  & 84\%  & 100\%& 100\%\\
    800Mpc& $10\mathrm{deg^2}$ & 94\% &  96\% & 94\%    & 51\%  & 98\% & 99\%\\
          & $5\mathrm{deg^2}$ & 84\% & 94\% & 79\%      & 13\%  & 96\% & 93\%\\
          & $1\mathrm{deg^2}$ & 40\%& 54\% & 35\%       & 0\%   & 64\% & 53\%\\
       \hline
           & $30\mathrm{deg^2}$ & 89\% & 95\% & 91\%    &34\%   & 98\% & 99\%\\
    1600Mpc& $10\mathrm{deg^2}$ &59\% & 78\% & 60\%     & 3\%   & 89\% & 79\%\\
           & $5\mathrm{deg^2}$ & 46\%& 59\% & 42\%      & 0\%   & 70\% & 58\%\\
           & $1\mathrm{deg^2}$ & 11\%& 14\% & 4\%       & 0\%   & 16\% & 9\%\\
      \hline
      \hline
  \end{tabular}
  \caption{A table showing the fraction of detectable \ac{BNS}s in  Simulation II which can be localized to within $\mathrm{30deg^2, 10deg^2, 5deg^2}$, and $1\text{deg}^2$ at the time of merger with tested networks (ETCN, CECN, C20N), ET-CE network (ETCE), and also ET-CE network together with a detector in Australia (1ET2CE, 2ET1CE).}\label{snr_er_number}
\end{table*} 
We calculated the localization uncertainty of binary neutron star (\ac{BNS}) mergers at the time of the merger using tested detector networks. This allowed us to compile percentages of detectable sources that could be localized within 30 deg$^2$, 10 deg$^2$, 5 deg$^2$, and 1 deg$^2$ for each detector network, providing a comprehensive perspective on their localization abilities. The results are summarized in Table~\ref{snr_er_number}. We observe that, across various distances and under the same maximum allowable localization uncertainty, the CECN network consistently detects the highest number of sources, with only a few exceptions. Additionally, the ETCN network generally outperforms the C20N network in terms of source detection. Comparing these results with networks incorporating a detector in Australia, we find that the 2ET1CE (1ET2CE) configuration tends to localize more sources than the ETCN (CECN) network within the same maximum allowable region. This suggests that integrating a detector in Australia into the ET-CE network could lead to more accurate localization (at least during the time of merger) for the same binary neutron star (\ac{BNS}) sources compared to adding a detector in China.\\
\indent Also, for three tested networks including the detector in China, we show the cumulative distribution of localization uncertainty at the time of merger for the detectable sources (\ac{SNR} no less than 8) at each fixed distance in Figure~\ref{er_appendix} of the appendix. Additionally, Table~\ref{loct_fixed} provides quantitative values of the corresponding localization uncertainty for the top 90\%, 50\%, and 10\% of best-localized sources by each network at fixed distances. 
For example, for the detectable sources at 40\,Mpc, ETCN could localize $90\%$ best-localized sources to within 0.019 deg$^2$. Take 200\,Mpc results in Table~\ref{loct_fixed} as an example, we can see that: (1) The \ac{ET}-\ac{CE} network can localize 90\% of the best-localized sources within 2.175 deg$^2$. When an additional detector is added in China to the \ac{ET}-\ac{CE} network, the localization error decreases to approximately 10\% to 22\% of its original value;(2) We observe that, for the top 90\% and 50\% of best-localized sources, the ETCN (CECN) network consistently exhibits larger localization uncertainty compared to the 2ET1CE (1ET2CE) network. However, for the top 10\% of best-localized sources, the ETCN (CECN) network tends to demonstrate comparable or even smaller localization uncertainty than the 2ET1CE (1ET2CE) network. These findings suggest that augmenting the \ac{ET}-\ac{CE} network with a detector in Australia generally results in overall smaller localization uncertainty for the majority of sources, while adding a detector in China may lead to better localization for a very small subset of sources compared to Australia; (3) In terms of comparing the three networks featuring detectors in China, an intuitive examination of Figure~\ref{er_appendix} reveals that the calculated localization uncertainty values by CECN are notably more concentrated on smaller values. Quantitatively, for the top 90\% best-localized sources, the localization uncertainty value for CECN is 0.220 deg$^2$, representing 45\% of the value of ETCN and 52\% of the value of C20N. This suggests that CECN exhibits a more accurate localization ability for the same binary neutron star (\ac{BNS}) sources, followed by C20N, and then ETCN.
\begin{table*}
  \centering
  \begin{tabular}{p{1.5cm}p{0.8cm}p{0.8cm}<{\centering}p{1cm}<{\centering}p{1cm}<{\centering}p{1cm}<{\centering}p{1.2cm}<{\centering}p{1.2cm}<{\centering}}
      \hline
      \hline
       &  & ETCN & CECN & C20N & ETCE & 1ET2CE & 2ET1CE\\
       \hline
                           & 90\% & $0.019$ & $0.009$ & $0.017$  & 0.087 & 0.007 & 0.010\\
       40Mpc               & 50\% & $0.004$ & $0.002$ & $0.004$  & 0.024 & 0.001 & 0.002\\
                           & 10\% & $0.001$& $0.001$ & $0.001$   & 0.010 & 0.001 & 0.001\\
      \hline
                           & 90\% & $0.486$ & $0.220$ & $0.427$  & 2.175 & 0.18 & 0.25\\
       200Mpc              & 50\% & $0.099$ & $0.055$ & $0.106$  & 0.604 & 0.04 & 0.06\\
                           & 10\% & $0.015$& $0.013$ & $0.024$   & 0.243 & 0.01 & 0.02\\
       \hline
                           & 90\% & $1.945$ & $0.880$ & $1.709$  & 8.701 & 0.72 & 1.02\\
       400Mpc              & 50\% & $0.396$ & $0.220$ & $0.422$  & 2.416 & 0.15 & 0.22\\
                           & 10\% &$0.060$& $0.050$ & $0.096$    & 0.971 & 0.06 & 0.06\\
       \hline
                           & 90\% & $7.782$& $ 3.521$ & $6.836$  & 34.803 & 2.89 & 4.07\\
       800Mpc              & 50\% & $1.584$ & $0.881$ & $1.690$  & 9.665  & 0.60 & 0.91\\
                           & 10\% & $0.241$& $0.200$ & $0.384$   & 3.885  & 0.23 & 0.26\\
       \hline
                           & 90\% & $24.150$ &$14.083$ & $27.345$ & 136.411 &10.07 & 14.65\\
      1600Mpc              & 50\% & $5.737$ & $3.524$ & $6.758$   & 36.735  &2.30  & 3.57\\
                           & 10\% & $0.931$& $0.801$ & $1.536$    & 15.318  &0.90  & 1.03\\
      \hline
      \hline
  \end{tabular}
  \caption{The 90\% credible regions of localization uncertainty in deg$^2$ for detectable \ac{BNS}s in  Simulation II. The detectable sources are selected according to \ac{SNR} larger than 8. Then in detectable sources, values for 90\%, 50\% and 10\% respectively correspond to 90\%, 50\% and 10\% of cumulative distribution, which mean localization uncertainties for best-localized 90\%, 50\% and 10\% of sources. From the third column to the last column, are respectively the results of ETCN network, CECN network, C20N network, ET-CE network (ETCE), ET-CE network together with a CE-like detector in Australia (1ET2CE), ET-CE network together with an ET-like detector in Australia (2ET1CE).}\label{loct_fixed}
  \end{table*} 
  \subsubsection{Early warning}  

\begin{figure*}[h]
\subfigure[$1\text{deg}^2$] {\includegraphics[width=1\columnwidth]{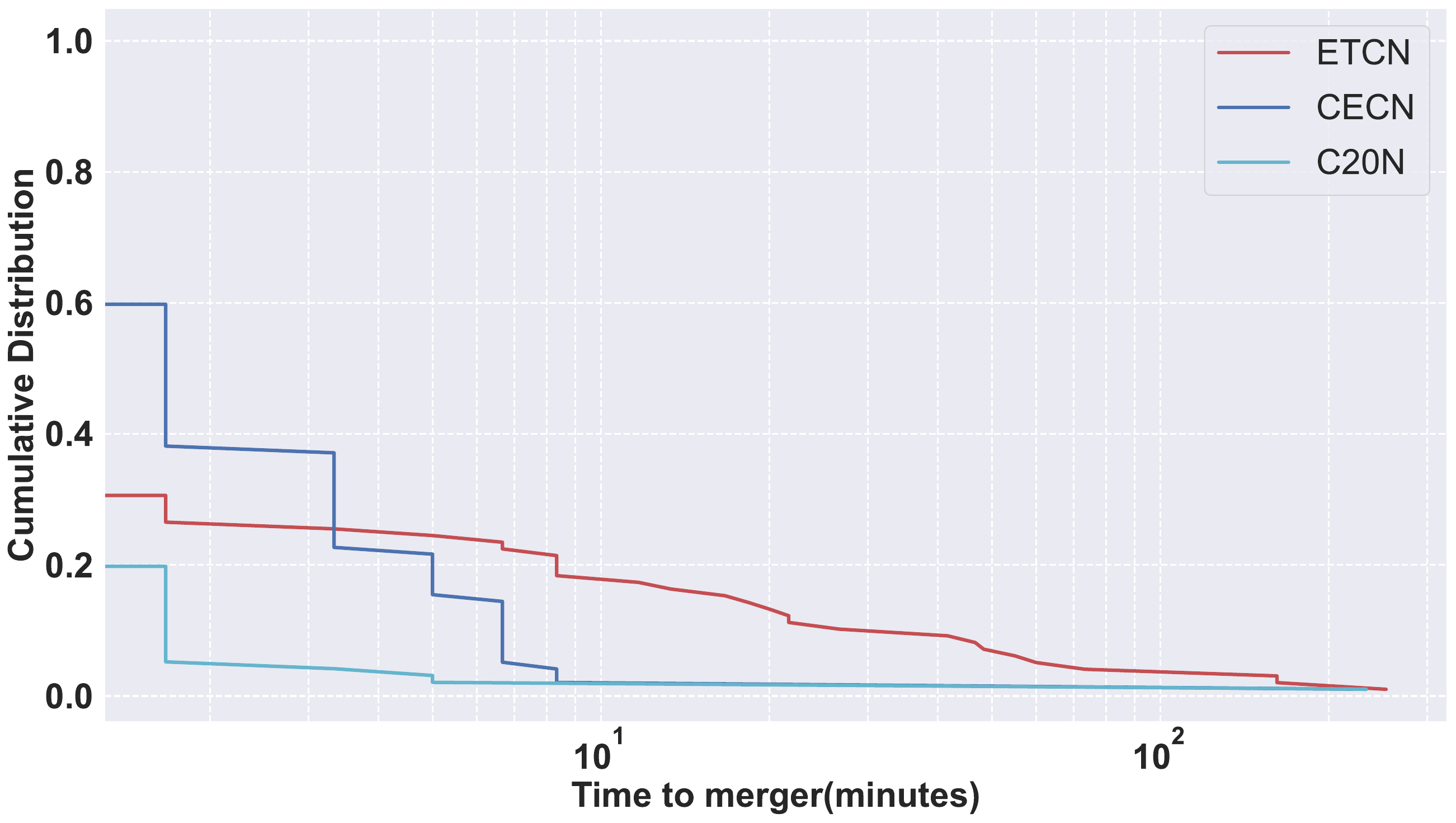}} 
\subfigure[$5\text{deg}^2$] {
\includegraphics[width=1\columnwidth]{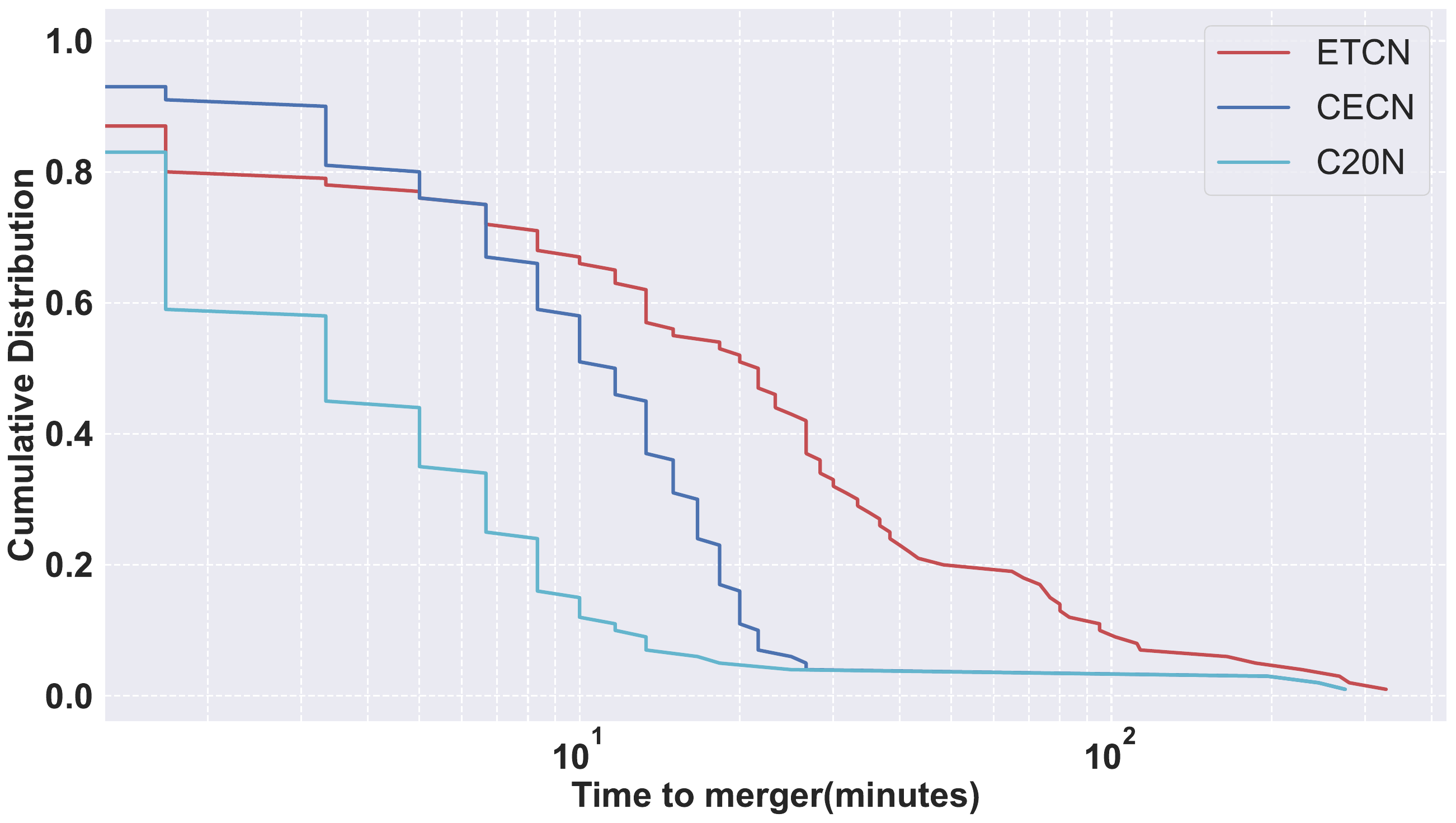}} 

\subfigure[$10\text{deg}^2$] {
\includegraphics[width=1\columnwidth]{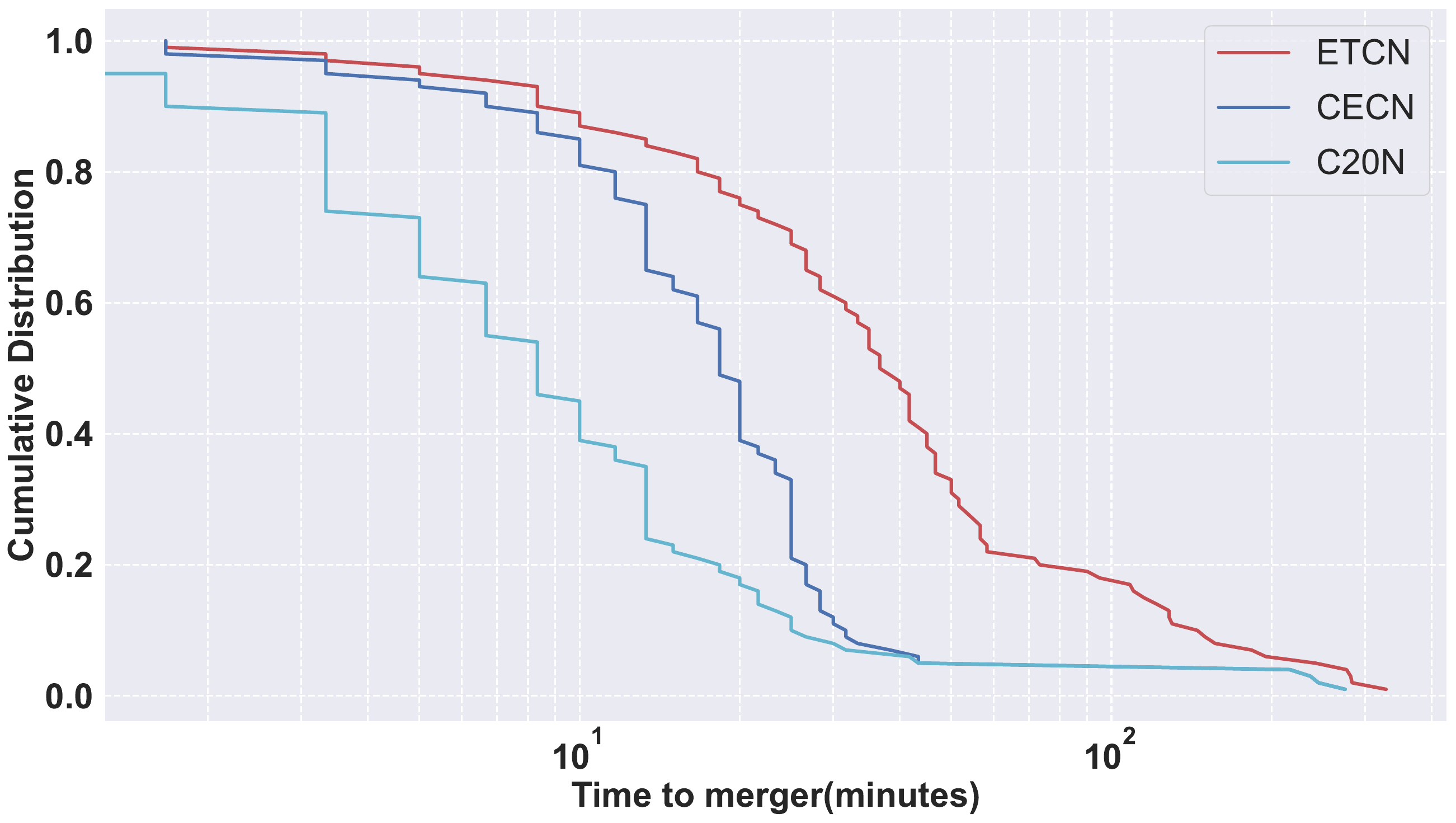}} 
\subfigure[$30\text{deg}^2$] {
\includegraphics[width=1\columnwidth]{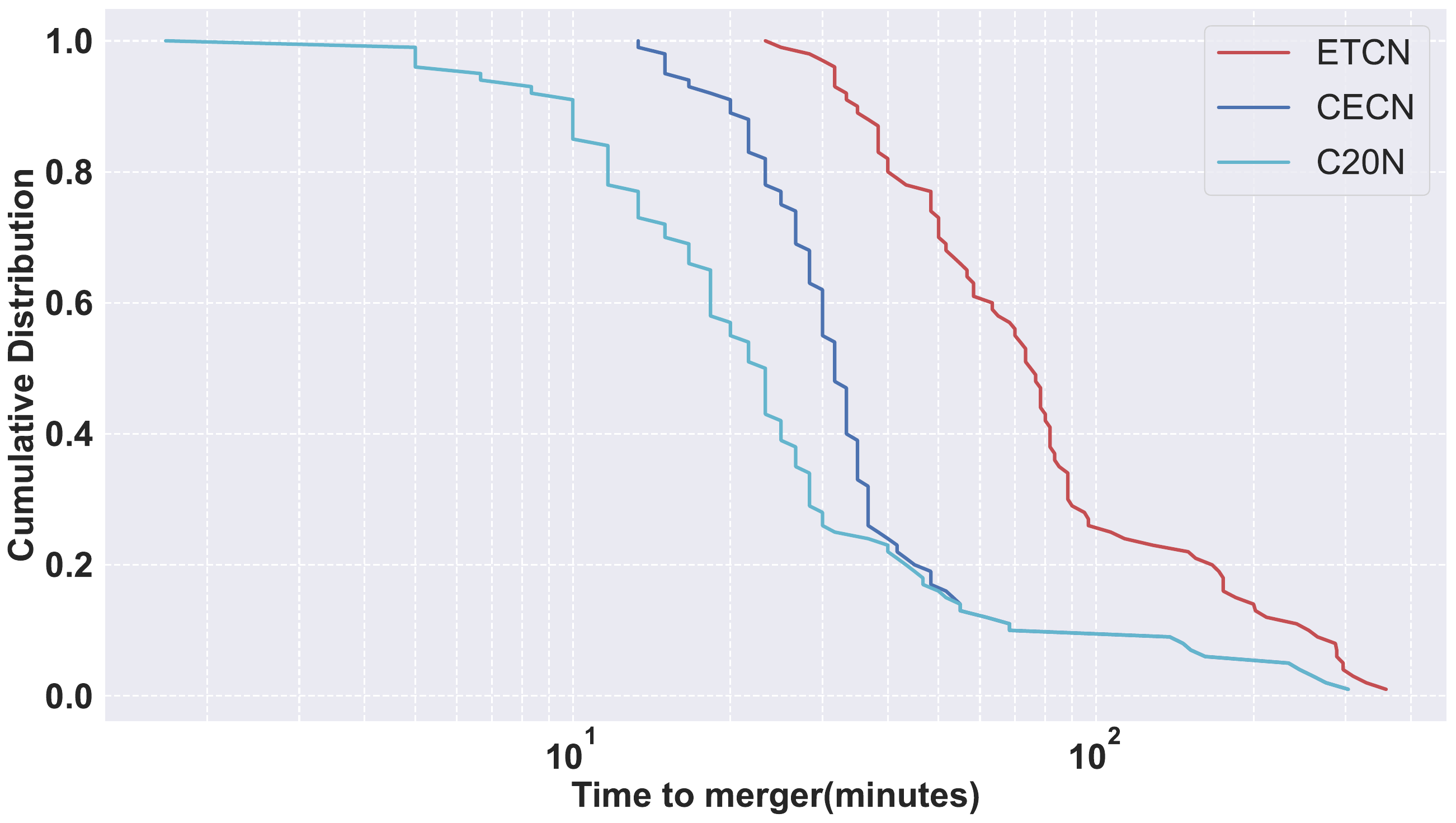}}
\caption{The cumulative distribution of time to merger for those \ac{BNS}s at 200 Mpc that achieve the early warning criteria. The x-axis is the time to merger in minutes when the signal meets the localization requirement denoted by deg$^2$ in subtitle. The y-axis is the cumulative distribution of detectable events that achieve these early warning criteria. All distributions are normalized such that the first bin equals 1.}\label{200EW}
\end{figure*}

  \indent To investigate the capability of tested detector networks for giving early warning alerts, we did some statistics, firstly, for a given sky area, we selected sources that meet two requirements of early warning criteria at the time of merger. Next, based on these selected sources, we obtained the earliest time when the source achieve the early warning criteria. Given respectively the maximum allowable sky area of 1deg$^2$, 5deg$^2$, 10deg$^2$, 30deg$^2$, the cumulative distribution of time to merger (obtained by time of merger - earliest time) when selected \ac{BNS}s at 200\,Mpc achieves the early warning criteria by tested detector networks are shown in Figure~\ref{200EW} (Similar figures for \ac{BNS}s at other fixed distances can be found in the Appendix. ), which shows that in general, the line of ETCN is shifted right the furthest towards higher values, indicating that it achieves the early warning criteria at the earliest time. Additionally, the overall slope of the line of ETCN is steeper than the other two, especially for a larger given sky area (e.g. 30 deg$^2$). Steeper slopes indicate a tighter range of values and, therefore, lower variability.\\
  \indent One method to read a cumulative distribution in a quantitative way is to look up percentiles. For example, for each maximum localization uncertainty case (e.g. 30 deg$^2$), we extract values of time to merger which correspond to respectively 90\%, 50\%, 10\% of for detectable sources at 200\,Mpc from Figure~\ref{200EW}, and the numbers are listed in Table~\ref{pre-merger-cumu} (the extracted values represent the lower bound of time to merger). Take ETCN and 30 deg$^2$ as an example, an early warning can be issued for 90\% of the sources as early as $\geq$35 minutes before merger. From this table, we can infer the following: (1) given 30 deg$^2$ as the maximum allowable region, for 90\% detectable sources, ET-CE network could give alerts at least 3.33 minutes before merger, adding a detector in China raise this lower limit to 10 minutes, and in some instances, alerts can be released as early as more than 300 minutes before merger; (2) with horizontal contrast with networks including detector in Australia, we can see that given the same maximum allowable sky area, the lower bound of time to merger of ETCN (CECN) is generally larger than 2ET1CE(1ET2CE) network, this means that, based on ET-CE network, adding a detector located in China tends to meet the early warning criteria and release an alert earlier before Australian detector network case; (3) for a relatively large given maximum allowable region, for example, 10 deg$^2$ and 30 deg$^2$, three detector networks including detector in China, from C20N, CECN, to ETCN, show an increasing trend of lower limit of time to merger, as intuitively shown in the Figure~\ref{200EW}, the three lines are arranged from bottom to top, which means ETCN could achieve the early warning criteria earlier than CECN, and CECN is earlier than C20N. While in 1 deg$^2$ case, for 50\% detectable sources, the lower limit of time to merger of ETCN and C20N network are both 0, and it is 1.67 minutes for CECN network. This means most detectable sources meet the early warning criteria nearly at the time of merger by ETCN and C20N, while at the same time, most detectable sources achieve the early warning criteria by CECN network before merger and are concentrated in a very narrow range (mainly $\sim$ 1.67-10 minutes before merger), this is because when \ac{BNS} is close to merger, the high sensitivity of CE in this frequency range (e.g., 10 Hz - 1000 Hz) enables its detected localization uncertainty to rapidly reach 1 $deg^2$ within a few minutes before the merger. And the sensitivity of ET and 20km detector is similar (not better than CE), so the localization uncertainty detected by these two detector networks tends to decrease more slowly to 1 deg$^2$, in this case, most detectable sources could be localized to 1 deg$^2$ only at the time of merger.\\
  \indent We also did another statistic for localization before merger, as seen in Table~\ref{pre-merger}. Taken 200\,Mpc, ETCN network and pre-merger time of 10 minutes as an example, we firstly counted the number of sources of which could be detected with SNR $\geq$ 8 and localized within 10 deg$^2$ at the time of merger, then for these selected sources, we counted the number of sources of which could be detected with SNR $\geq$ 8 and localized within 100 deg$^2$ at 10 minutes before the merger. Finally, the ratio between these two counts are calculated and the values are shown in Table~\ref{pre-merger} as percentages. For \ac{BNS}s at one selected distance, e.g., 400\,Mpc, and by one selected network, we will see that the number corresponding to pre-merger time of 10 min, 20 min, and 30 min is decreasing, which is easy to understand, because the denominator stays the same, and the localization uncertainty is decreasing as time approaches the merger, which means that the localization uncertainty corresponding to 10 min, 20 min, and 30 min is increasing, therefore, the number of sources which meet the criteria in numerator is decreasing. And we can infer the detector network sensitivity from its decreasing rate from 10 min case to 30 min case, usually, when the detector network has better sensitivity during this range, its detected localization uncertainty will decrease quickly with time, thus resulting in more precise source localization. We could find that CECN decreases more quickly than ETCN network, which is because, CECN has better sensitivity especially in 10 Hz - 1000 Hz. \\
\begin{table}
\centering
  \begin{tabular}{p{0.9cm}p{0.5cm}cccccc}
      \hline
      \hline
                                && ETCN  & CECN   &  C20N & ETCE & 1ET2CE & 2ET1CE\\
       \hline
                              &90\%   & 35    & 20     & 10    & 3.33 &  16.67& 18.33\\
         $30\mathrm{deg^2}$   &50\%   & 75    & 31.67  & 23.33 & 18.33 & 28.33& 45\\
                              &10\%   & 255   & 68.33  & 68.33 & 68.33 & 68.33& 75\\
           \hline
                              &90\%   & 8.33  & 6.67  & 1.67   & 0    &6.67 &6.67\\
   $10\mathrm{deg^2}$         &50\%   & 36.67 & 18.33 & 8.33   & 3.33 &15 &28.33\\
                              &10\%   & 145   & 31.67 & 25     & 25 & 30 & 65\\
            \hline
                              &90\%   & 0     & 3.33  & 0      & 0    & 3.33 & 1.67\\
         $5\mathrm{deg^2}$    &50\%   & 21.67 & 11.67 & 3.33   & 0    & 8.33 & 13.33\\
                              &10\%   & 95    & 21.67 & 11.67  & 6.67 & 18.33 & 48.33\\
            \hline                  
                              &90\%   & 0     & 0     & 0      & 0   & 0 &0\\
         $1\mathrm{deg^2}$    &50\%   & 0     & 1.67  & 0      & 0   & 1.67 & 0\\
                              &10\%   & 29.67 & 6.67  & 1.67   & 0   & 5 & 11.67\\
      \hline
      \hline
  \end{tabular}
  \caption{A table showing that given the maximum allowable region of respectively $\mathrm{30deg^2, 10deg^2, 5deg^2}$, and $1\text{deg}^2$, the time to the merger in minutes,  90\%, 50\%, 10\% of the cumulative distribution of time to the merger of detectable events at 200 Mpc by each network. }\label{pre-merger-cumu}
\end{table} 
\begin{table*}
  \centering
  \begin{tabular}{p{1.5cm}p{1.3cm}p{0.9cm}p{0.9cm}p{0.9cm}p{1.3cm}p{1.3cm}p{1.3cm}}
      \hline
      \hline
          & pre-merger time (min) & ETCN & CECN & C20N & ETCE & 1ET2CE & 2ET1CE\\
       \hline
                           & 10 & 100\% & 100\% & 100\% & 100\% & 100\% & 100\%\\
       40Mpc               & 20 & 100\% & 100\% & 100\% & 100\% & 100\% & 100\%\\
                           & 30 & 100\% & 100\% & 100\% & 100\% & 100\% & 100\%\\
      \hline
                           & 10 & 100\% & 100\% & 100\% & 100\% & 100\% & 93\%\\
       200Mpc              & 20 & 100\% & 100\% & 100\% & 99\%  & 99\%  & 91\%\\
                           & 30 & 100\% & 99\%  & 81\%  & 78.8\%& 92\%  & 80\%\\
       \hline
                           & 10 & 100\% & 100\% & 91\%  & 75\%  & 97\% & 48\%\\
       400Mpc              & 20 & 100\% & 90\%  & 57\%  & 46.7\%& 75\% & 32\%\\
                           & 30 & 94\%  & 62\%  & 27\%  & 28.3\%& 34\% & 25\%\\
       \hline
                           & 10 & 81\% & 77.1\% & 38.3\% & 29.4\%& 67.3\% & 5.3\%\\
       800Mpc              & 20 & 66\% & 35.4\% & 6.4\%  & 11.8\%& 15.3\% & 0\\
                           & 30 & 50\% & 6.3\%  & 5.3\%  & 9.8\% & 1.0\%  & 0\\
       \hline
                           & 10 & 44.1\%& 25.6\%& 3.33\% & 66.67\% &9.0\% & 0\\
      1600Mpc              & 20 & 30.5\%& 2.6\% & 3.33\% & 66.67\% &0\%  & 0\\
                           & 30 & 25.4\%& 2.6\% & 3.33\% & 66.67\% &0\%  & 0\\
      \hline
      \hline
  \end{tabular}
  \caption{Pre-merger localization, number in the table represents $N_{\Delta \Omega<100 \mathrm{deg}^2, \mathrm{t}_{\mathrm{pre}}} / N_{\Delta \Omega<10 \mathrm{deg}^2, \mathrm{t}_{\text {merger }}}$. The SNR threshold is set to be 8. }\label{pre-merger}
  \end{table*} 
\section{Conclusion and Discussion}\label{discussion}
In this work, we use Fisher matrix method to estimate the detection rate, localization uncertainty and early warning performance of three test networks containing detector in China (ETCN, CECN, C20N) for simulated \ac{BNS} merger events (two kinds of population: Simulation I, \ac{BNS} mergers following the delay time distribution; Simulation II, \ac{BNS} mergers at fixed distances).The results were compared with those obtained from the ET-CE network, and with those obtained from ET-CE plus a detector located in Australia (2ET1CE: ET-CE network + an ET-like detector in Australia; 1ET2CE: ET-CE network + a CE-like detector in Australia).\\
\indent Firstly, for Simulation I (500 \ac{BNS} mergers that follow the delay-time distribution with redshift less than 2), using 8 as the network SNR threshold, the ET-CE network can detect 87\% of the sources, and could localize the best-localized 90\% sources (The following mentioned localization uncertainties are all for best localized 90\% sources) within 881 deg$^2$. Building upon the \ac{ET}-\ac{CE} network, the addition of a detector situated in China enhances the detection capability by at least 4.4\% and concurrently reduces localization uncertainty to less than 20\% of its original value, reaching within 162 deg$^2$. Contrasting these outcomes with the results of incorporating an Australian detector into the \ac{ET}-\ac{CE} network, we observe that when an ET-like detector is added to the \ac{ET}-\ac{CE} network, placing the detector in China (ETCN) yields a 1.6\% increase in detected sources compared to placement in Australia (2ET1CE). However, the localization error also escalates to approximately 140\% of the latter scenario. Conversely, with the addition of a CE-like detector to the \ac{ET}-\ac{CE} network, placing the detector in China (CECN) results in a 1\% increase in detected sources compared to placement in Australia (1ET2CE), with a localization error of approximately 154\% of the latter.\\
\indent Among the three networks containing detectors in China, CECN boasts the highest detection rate, reaching up to 98\%, and exhibits the smallest localization error ($\leq$87.43 deg$^2$). Conversely, in the other two networks, C20N achieves a 0.8\% higher detection rate than ETCN, albeit with a larger localization uncertainty, approximately 125\% of ETCN's. When considering a maximum allowable sky area, such as 30 deg$^2$, the proportions of sources detectable with \ac{SNR} $\geq$ 8 and localized within 30 deg$^2$ at the time of merger are 54\%, 67\%, and 49\% for ETCN, CECN, and C20N, respectively. Moreover, for these selected sources, after identifying all instances where they meet the criteria and their corresponding localization uncertainties, we observed that the time distribution of CECN and C20N is similar. In these two networks, the majority of sources reach a localization uncertainty within 30 deg$^2$ within 1 hour. In contrast, for the ETCN network, most sources are distributed within 2 hours. This indicates that compared with CECN and C20N, the ETCN network accumulates a \ac{SNR} greater than 8 at an earlier time before the merger, thereby improving the localization accuracy to within 30 deg$^2$.\\
\indent Furthermore, in Simulation II (which focuses on \ac{BNS} mergers at fixed distances), we simulated 100 sources for each distance. As an illustration, considering a distance of 200 Mpc, the \ac{ET}-\ac{CE} network achieves localization of the top 90\% best-localized sources within 2.175 deg$^2$. Additionally, if the maximum allowable sky area is restricted to 30 deg$^2$, the \ac{ET}-\ac{CE} network can issue an alert at least 3.3 minutes before the merger for 90\% of the detected sources. Building upon the \ac{ET}-\ac{CE} network, the addition of a detector located in China leads to a substantial reduction in localization uncertainty, ranging from 10\% to 22\% of the value observed in the \ac{ET}-\ac{CE} network. Moreover, alerts can be issued at least 10 minutes before the merger event, and for select well-localized sources, alerts could even be released more than 300 minutes before the merger. Contrasting these outcomes with the results obtained from augmenting the \ac{ET}-\ac{CE} network with an Australian detector, we observe that when an ET-like detector is added to the \ac{ET}-\ac{CE} network, the localization error incurred by placing it in China (ETCN) is approximately 194\% of that experienced when placing it in Australia (2ET1CE). However, in terms of early warning, ETCN generally tends to issue alerts earlier than 2ET1CE in most cases. For instance, with a maximum allowable sky area of 30 deg$^2$, ETCN can provide alerts at least 35 minutes before the merger for 90\% of detections, compared to 18.33 minutes for 2ET1CE. Similarly, when a CE-like detector is added to the \ac{ET}-\ac{CE} network, the positioning error of CECN is approximately 122\% of that of 1ET2CE, and CECN tends to release alerts earlier than 1ET2CE. Therefore, it suggests that when detectors of the same configuration are added to the \ac{ET}-\ac{CE} network, positioning them in Australia could result in more accurate localization, whereas positioning them in China increases the likelihood of issuing an alert earlier. Among the three networks featuring detectors in China, CECN demonstrates the highest localization accuracy, successfully localizing 90\% of the best-localized sources at a distance of 200 Mpc within 0.22 deg$^2$. In comparison, ETCN exhibits a localization error that is 220\% of that of CECN, while C20N's error is 194\% of that of CECN. In terms of early warning, ETCN typically fulfills the early warning conditions earliest, followed by CECN, and then C20N. This order aligns with the sensitivity levels of the three detectors, as the sensitivity of ET is relatively higher in the frequency range of 1 Hz to 10 Hz. Consequently, \ac{SNR} accumulation can begin earlier with ET, leading to a gradual reduction in localization uncertainty. Conversely, the sensitivity of CE and C20 experiences a notable increase from approximately 10 Hz onwards. Consequently, \ac{SNR} accumulation and improvements in localization accuracy commence around this frequency range. Within the 10 Hz to 1000 Hz range, CE exhibits exceptionally high sensitivity, enabling rapid \ac{SNR} accumulation and reduction in localization error. This facilitates the attainment of the highest localization accuracy at the time of merger.\\
\indent In conclusion, within the context of third-generation detectors, such as the proposed ET and CE, the addition of a third detector resembling ET or CE can markedly enhance the localization accuracy of \ac{BNS} mergers. This enhancement can potentially reduce the error to as low as 20\% of that observed with ET-CE. Moreover, such additions enable earlier alert releases, a crucial factor for planning subsequent joint GW-EM observations. Integrating an ET-like or CE-like detector into the ET-CE network, when compared to placement in Australia, generally leads to a slight increase in localization error but often results in a modest improvement in detection rate and earlier alert release. Furthermore, among the three networks featuring detectors in China, CECN exhibits the highest detection rate and localization accuracy. However, C20N and ETCN each possess distinct advantages in localization performance for two kinds of simulated \ac{BNS} mergers. For early warning, ETCN can achieve the criteria earlier thus to release an alert earlier, then followed by CECN, and finally C20N. These results are of great significance for the future cooperative planning of global detectors. In addition, the evaluation of ET-CE plus a detector placed in China in terms of detection rate, localization ability and early warning performance is also of certain reference significance for whether to establish ground-based detector in China in the future.\\
\acknowledgments
We are grateful to Prof. Mohammadtaher Safarzadeh and Prof. Seyed Alireza Mortazavi for the helpful discussion about the binary neutron star merger rate. We are also grateful for computational resources provided by Cardiff University and funded by an STFC grant supporting UK Involvement in the Operation of Advanced LIGO.  This work was supported by the National Natural Science Foundation of China under grant Nos. 11922303, 11920101003. Xilong Fan is supported by  the Fundamental Research Funds for the Central Universities (No.2042022kf1182).  Y. L is supported by the National Program on Key Research and Development Project through grant No. 2016YFA0400804, and by the National Natural Science Foundation of China with grant No. Y913041V01, and by the Strategic Priority Research Program of the Chinese Academy of Sciences through grant No. XDB23040100. ISH is supported by the Science and Technology Research Council (grant No. ST/V005634/1). LJG is supported by National Key R\&D Program of China (No. 2023YFA1607902) and NSFC (No. 12273058). 
\newpage
\bibliographystyle{plain}
\bibliography{ref}
\newpage
\appendix
\section{Some extra figures}
\begin{figure*}
\centerline{\includegraphics[scale = 0.21]{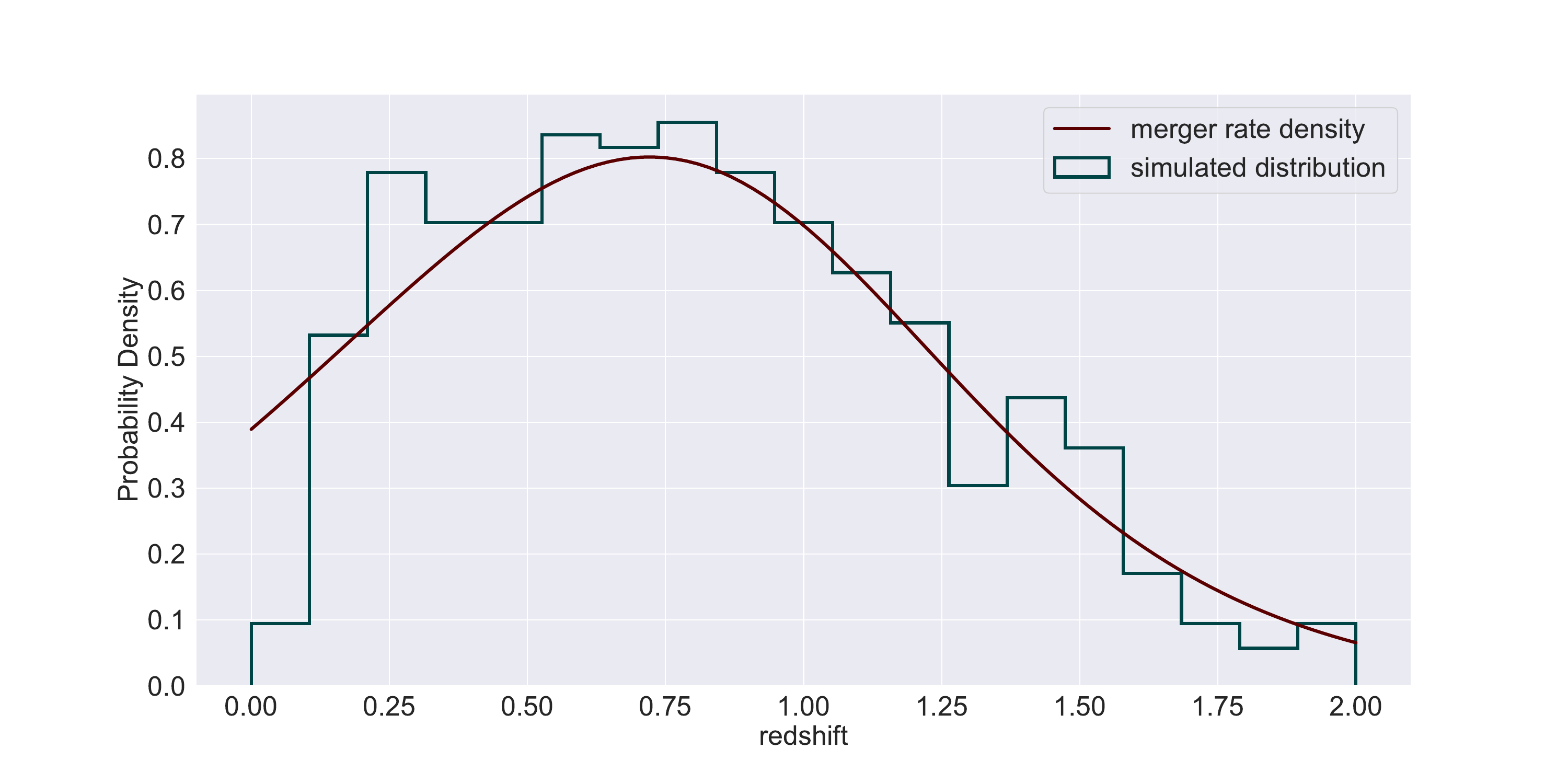}}
\caption{The distribution of the redshifts of the population of simulated \ac{BNS} mergers following the delay time distribution. The x-axis represents the redshift and the y-axis represents the probability density. The distribution is normalized such that the area under the curve is 1. The dark brown line is the merger rate density obtained from [\onlinecite{Safarzadeh2019b}] with $t_\text{{min}} = 1 \text{Gyr}, \Gamma = -1.5$,  and the blue line is the histogram distribution of simulated sources used for later analysis obtained by sampling method from merger rate density.}\label{z_distribution}
\end{figure*}
\begin{table*}
  \begin{tabular}{p{2.0cm}|c|c}
      \hline
      \hline
        & Simulation I: \ac{BNS} population & Simulation II: \ac{BNS} at fixed distances \\
        \hline

       \multirow{2}{*}{source number} & 500 sources & 100 sources respectively at\\
        & \multirow{2}{*}{} & 40 Mpc, 200 Mpc, 400 Mpc, \\
        & & 800 Mpc, 1600 Mpc \\
       \hline
       source~mass \quad\quad (in local frame)& 1.4 M$_{\odot}$ - 1.4 M$_{\odot}$ &  1.4 M$_{\odot}$ - 1.4 M$_{\odot}$\\  
       \hline
       waveform model & TaylorT3 & TaylorT3 \\  
      \hline
      signal duration &  $\sim$130 hours & $\sim$130 hours\\
      \hline
      redshift  &  follow delay time distribution with maximum $\leq 2$ & fixed \\
      \hline
      sky location &  randomized      &  randomized \\
      \hline
      inclination angle (cosine) &  randomized between (-1,1) &   randomized between (-1,1)\\
      \hline 
      polarization angle &  randomized between (0, 2$\pi$)      &   randomized between (0, 2$\pi$)  \\
      \hline
       \hline
  \end{tabular}
  \caption{Simulated source information. The second column shows the information for simulated \ac{BNS}s following the delay time distribution shown in Fig. \ref{z_distribution} (Simulation I). The last column shows the information for simulated \ac{BNS}s at fixed distances (Simulation II). }\label{source_info_table}
  \end{table*} 
\begin{figure*}
\centerline{\includegraphics[width=15cm]{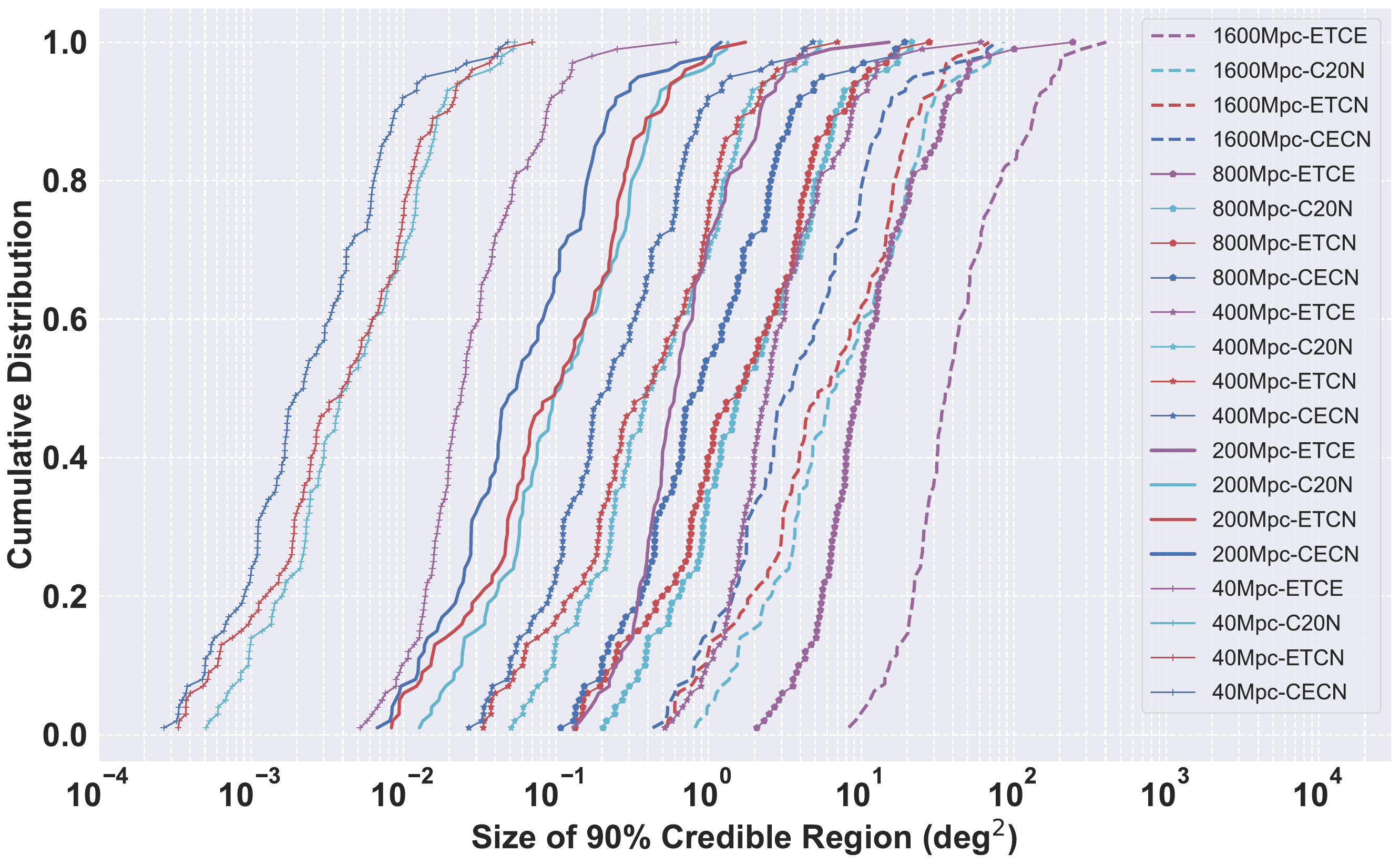}}
\caption{The cumulative distribution for localization uncertainty of \ac{BNS} sources at fixed distances.}\label{er_appendix}
\end{figure*}

\begin{figure*}
\subfigure[$1\text{deg}^2$] {
\includegraphics[width=1\columnwidth]{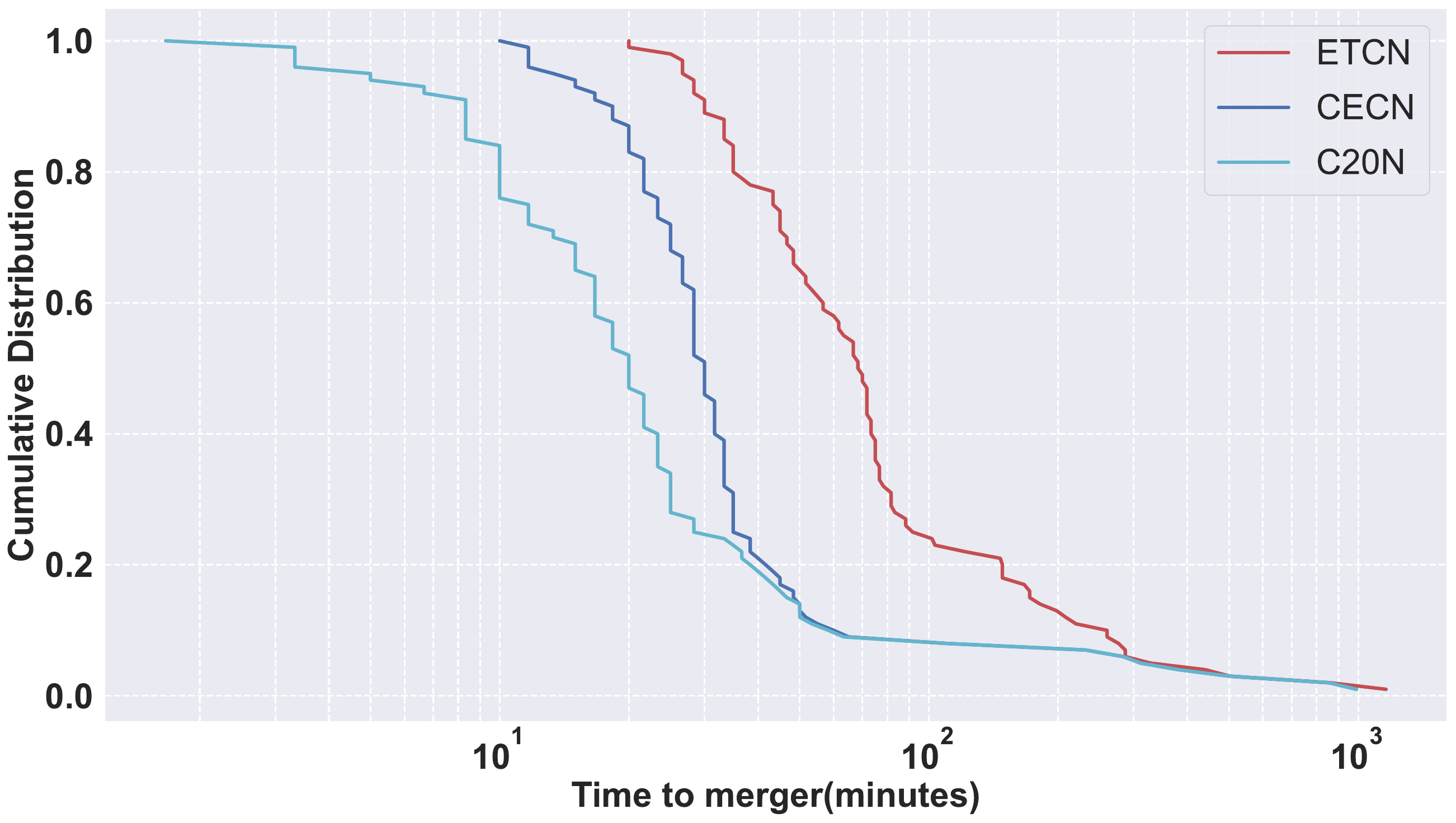}} 
\subfigure[$5\text{deg}^2$] {
\includegraphics[width=1\columnwidth]{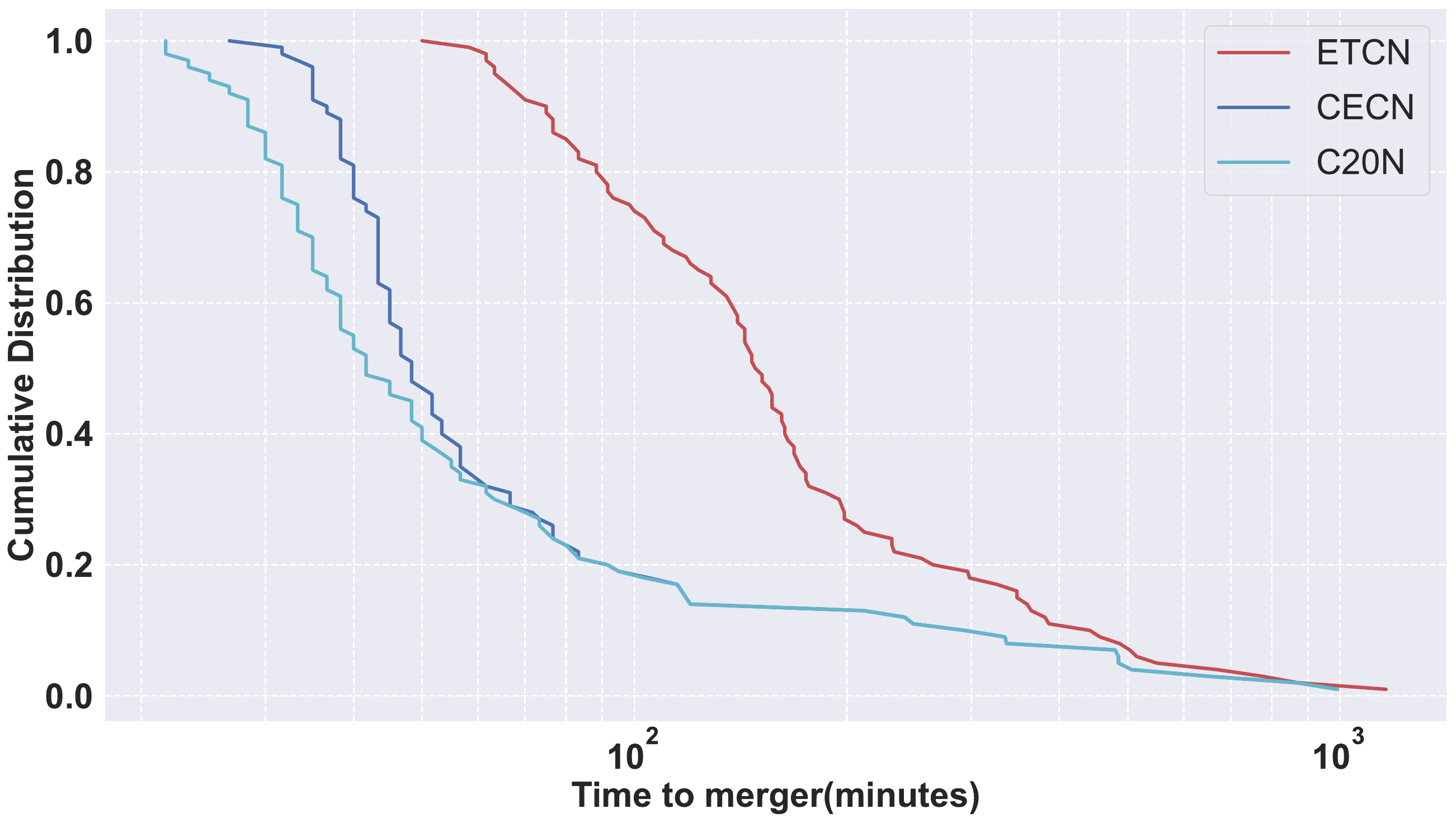}} 
\subfigure[$10\text{deg}^2$] {
\includegraphics[width=1\columnwidth]{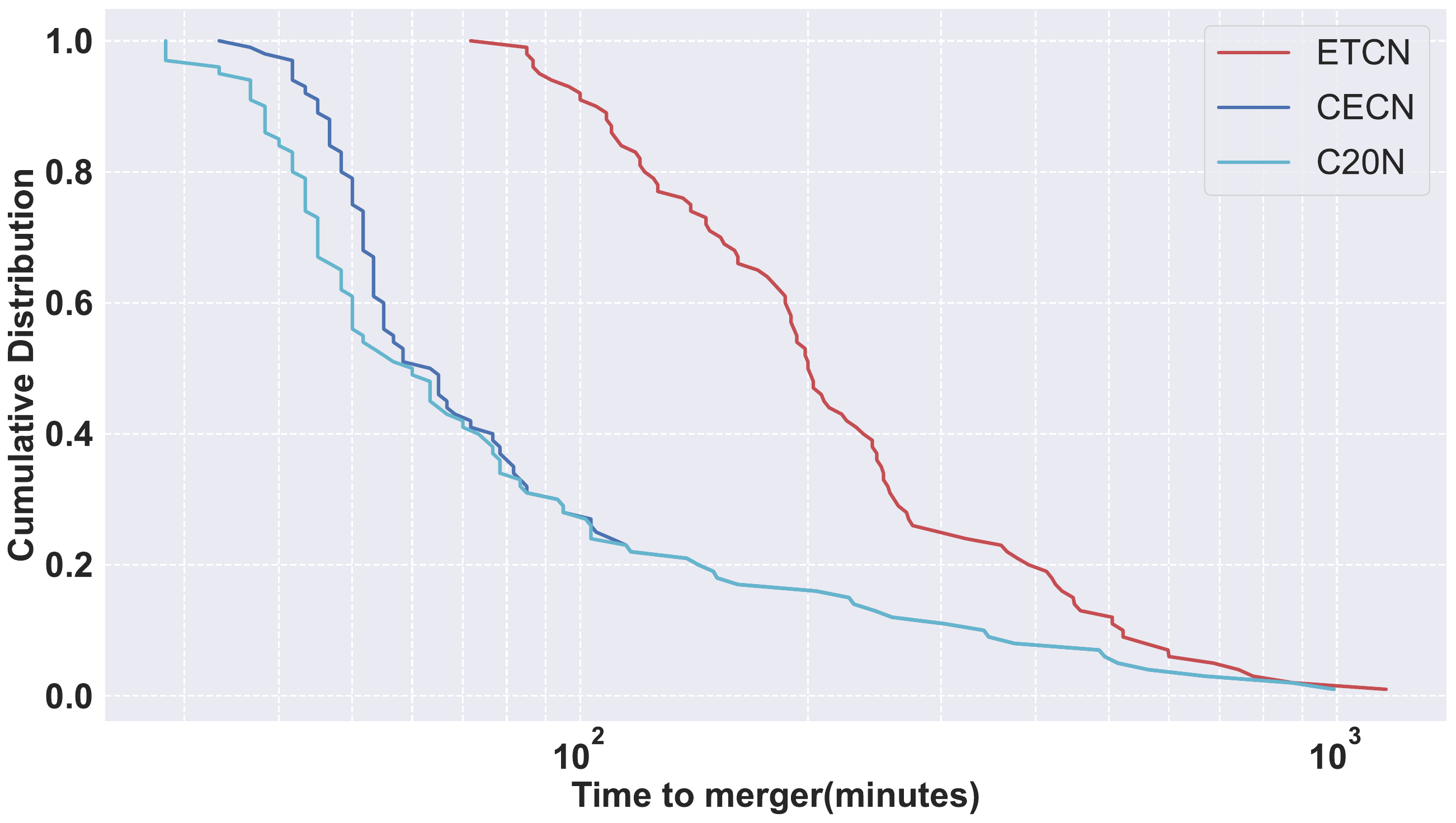}} 
\subfigure[$30\text{deg}^2$] { 
\includegraphics[width=1\columnwidth]{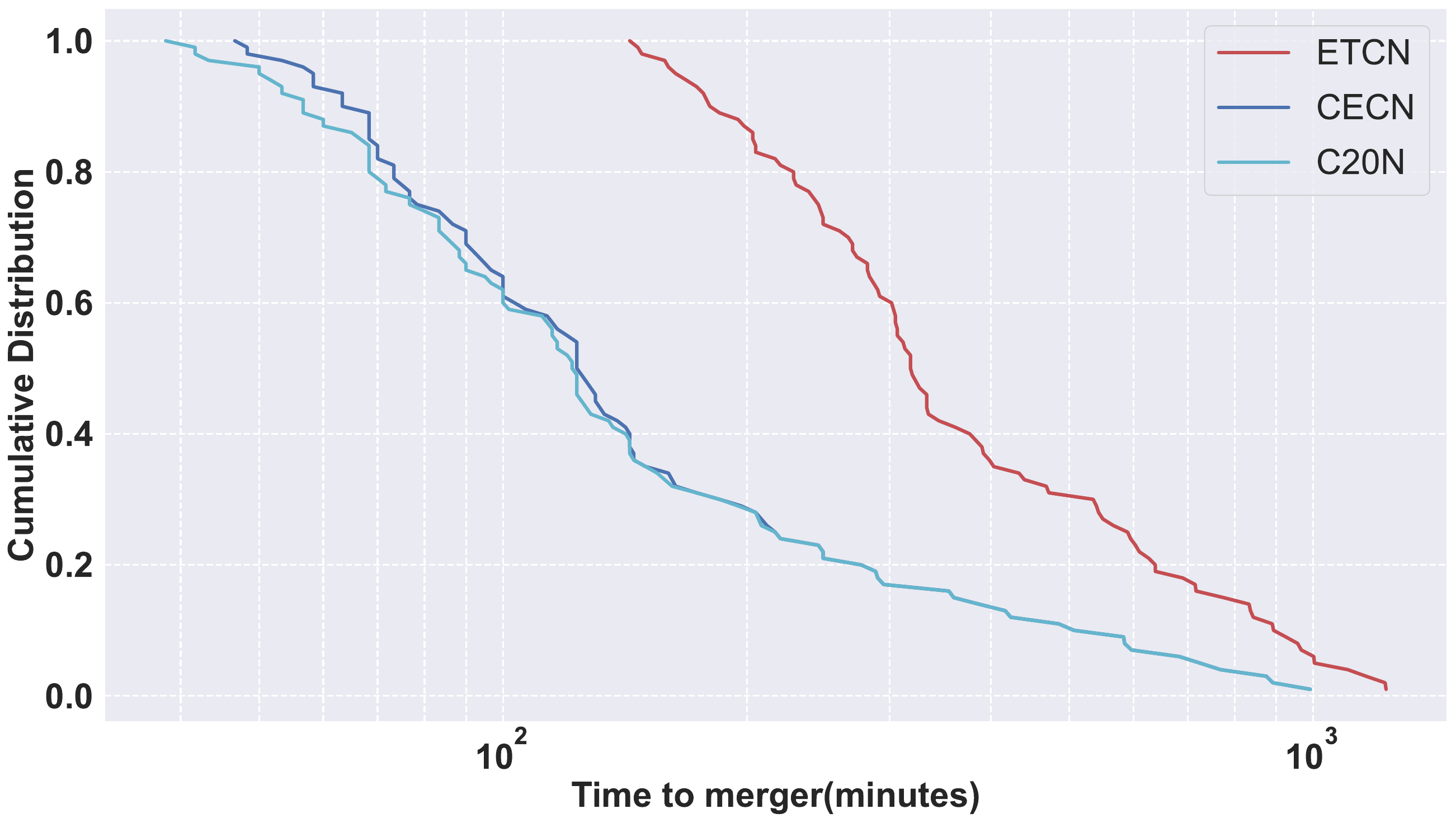}} 
\caption{The cumulative distribution of time to merger for \ac{BNS}s at 40 Mpc.}\label{40EW}
\end{figure*}

\begin{figure*}
\subfigure[$1\text{deg}^2$] { 
\includegraphics[width=1\columnwidth]{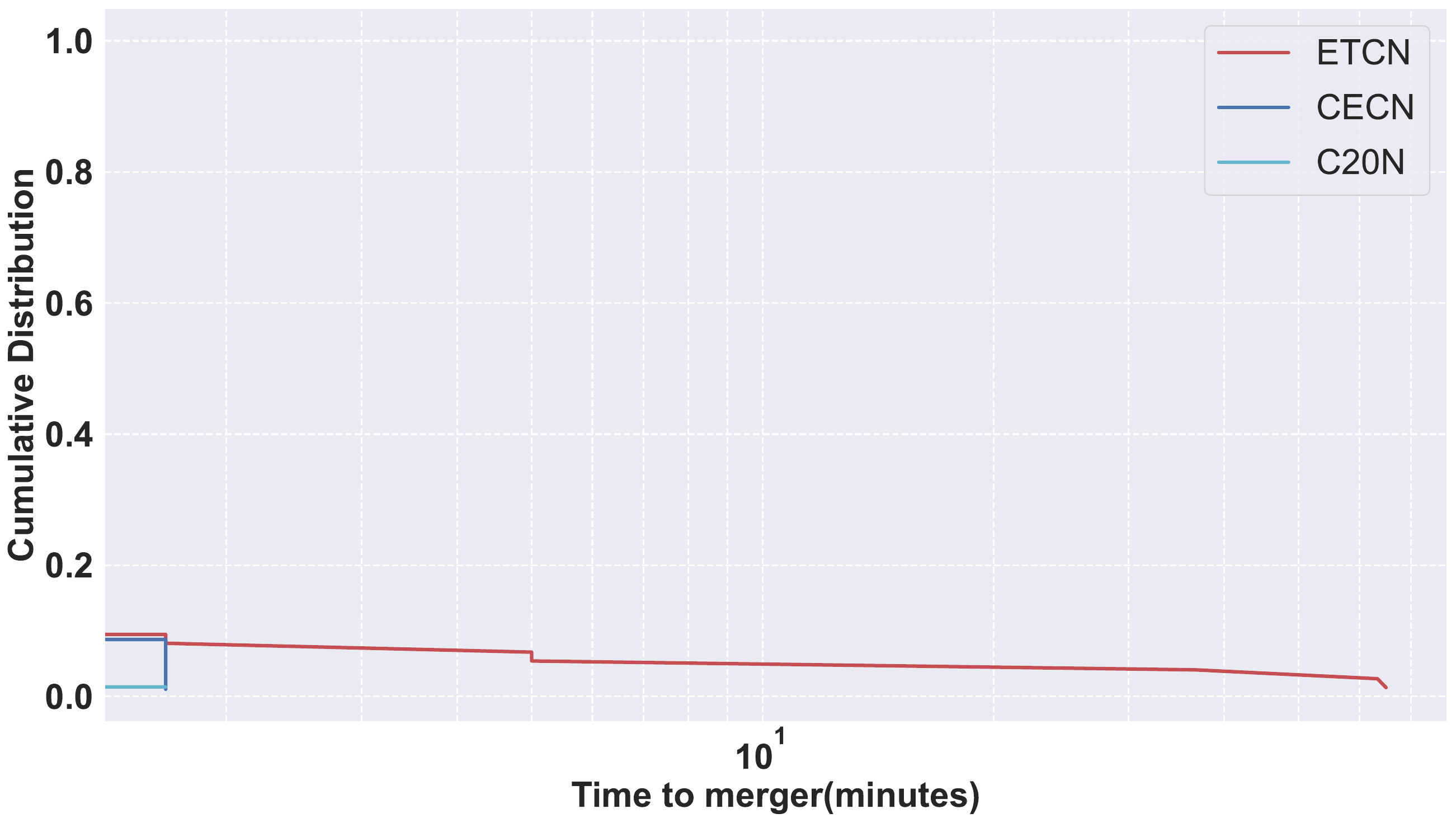}} 
\subfigure[$5\text{deg}^2$] {
\includegraphics[width=1\columnwidth]{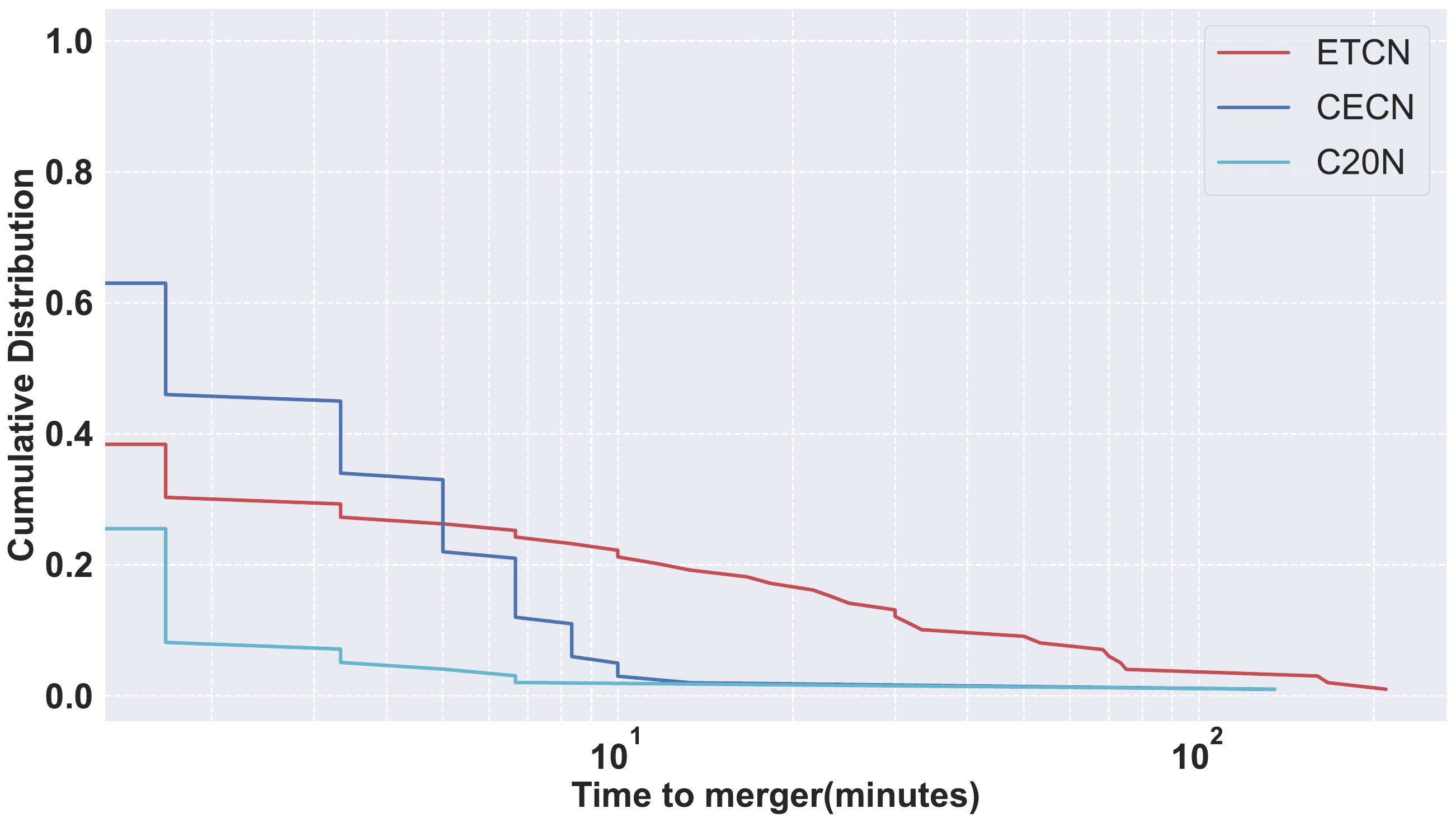}} 
\subfigure[$10\text{deg}^2$] {
\includegraphics[width=1\columnwidth]{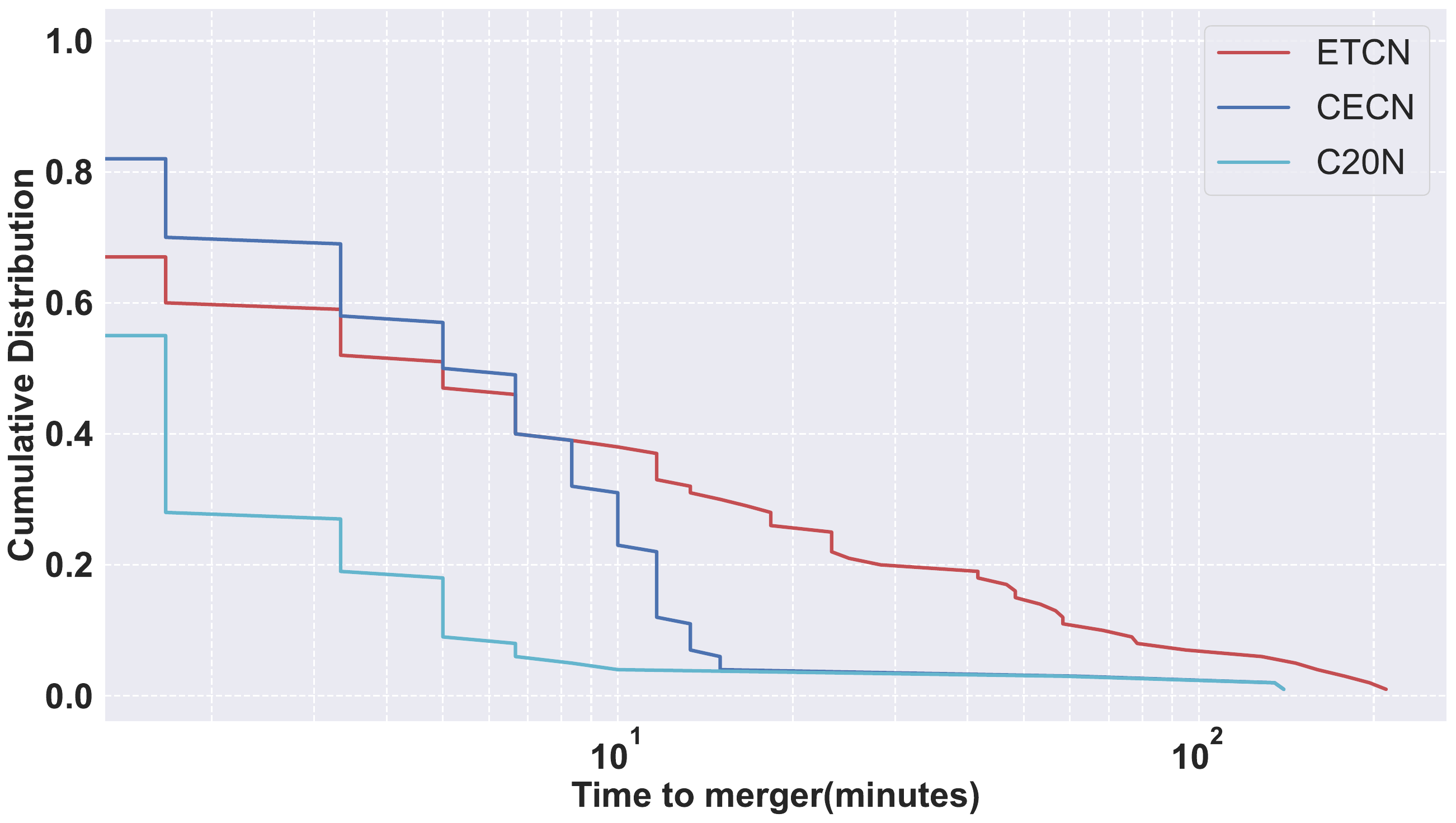}} 
\subfigure[$30\text{deg}^2$] { 
\includegraphics[width=1\columnwidth]{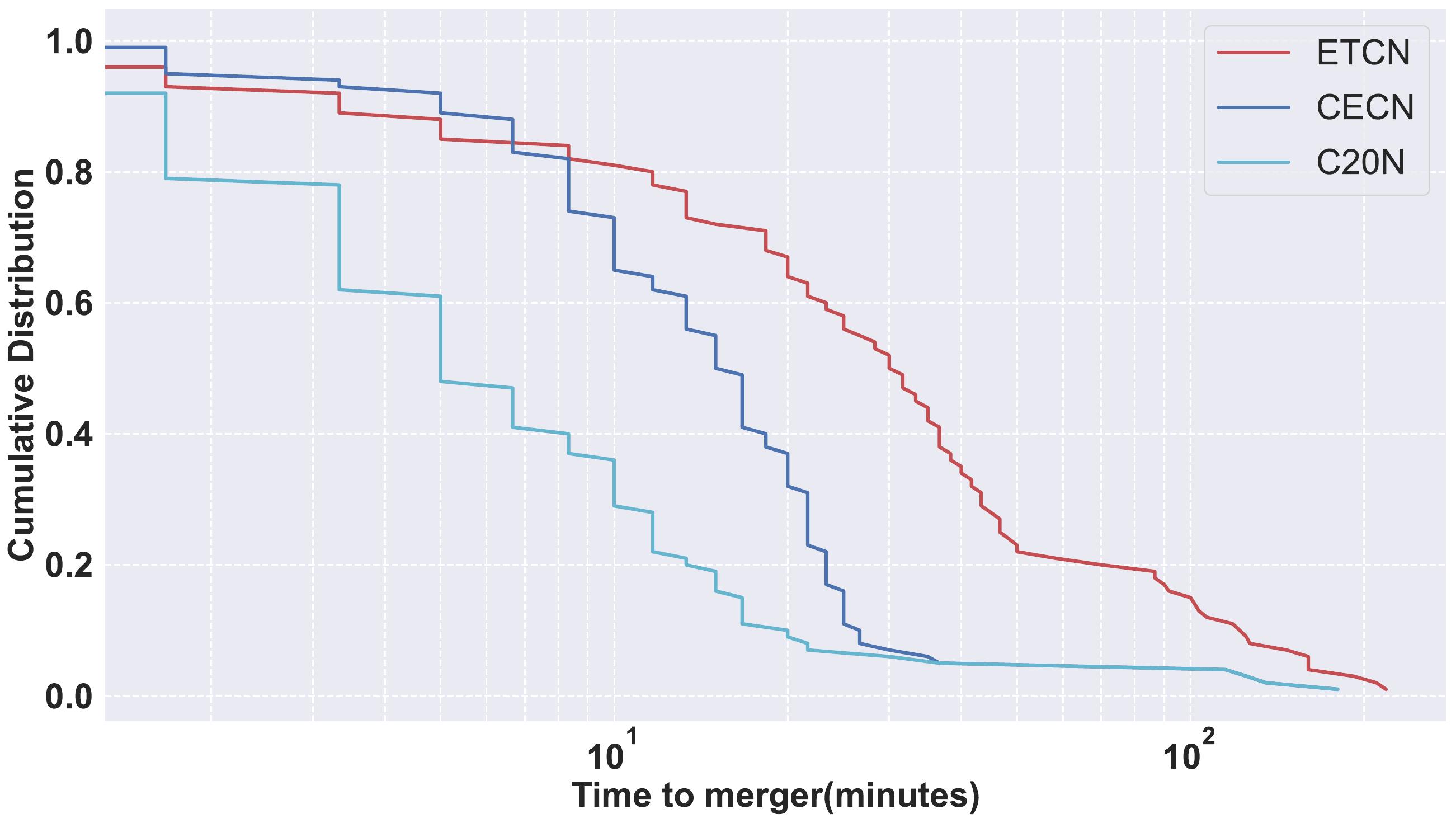}} 
\caption{The cumulative distribution of time to merger for \ac{BNS}s at 400 Mpc.}\label{400EW}
\end{figure*}

\begin{figure}
\subfigure[$10\text{deg}^2$] { 
\includegraphics[width=1\columnwidth]{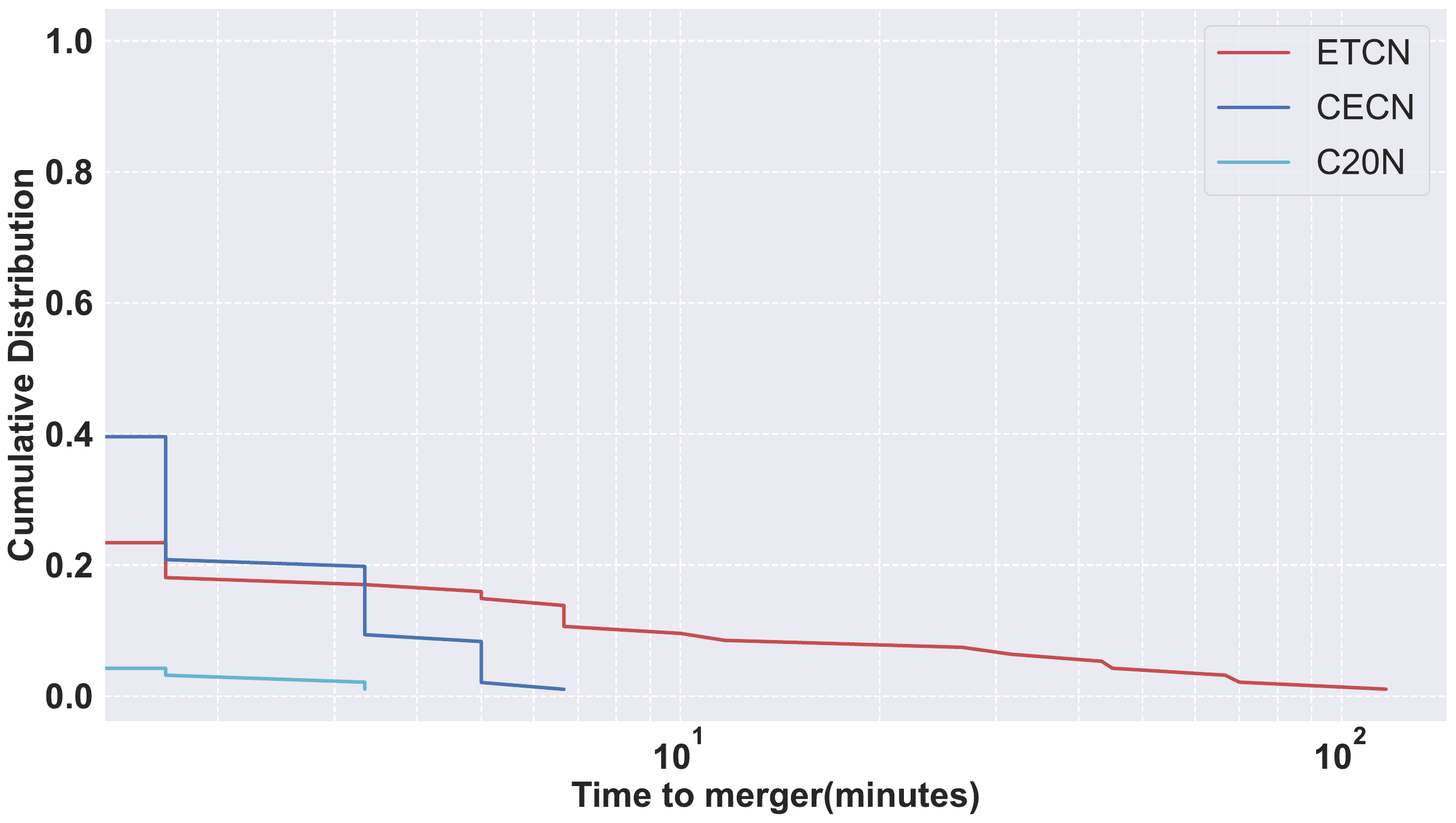}} 
\subfigure[$30\text{deg}^2$] {\includegraphics[width=1\columnwidth]{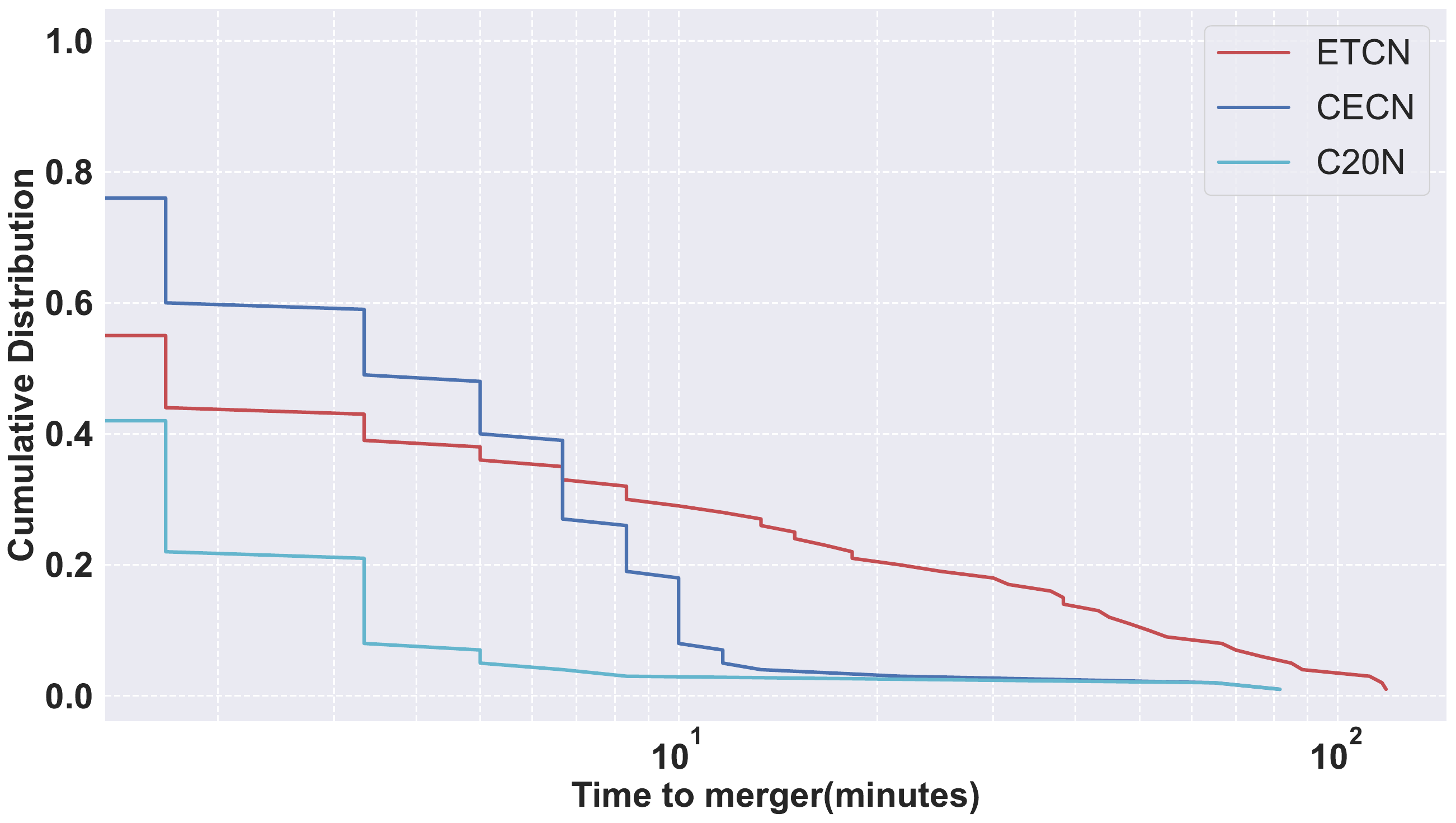}} 
\caption{The cumulative distribution of time to merger for \ac{BNS}s at 800 Mpc. }\label{800EW}
\end{figure}

\begin{figure}
\subfigure[$30\text{deg}^2$] {\includegraphics[width=1\columnwidth]{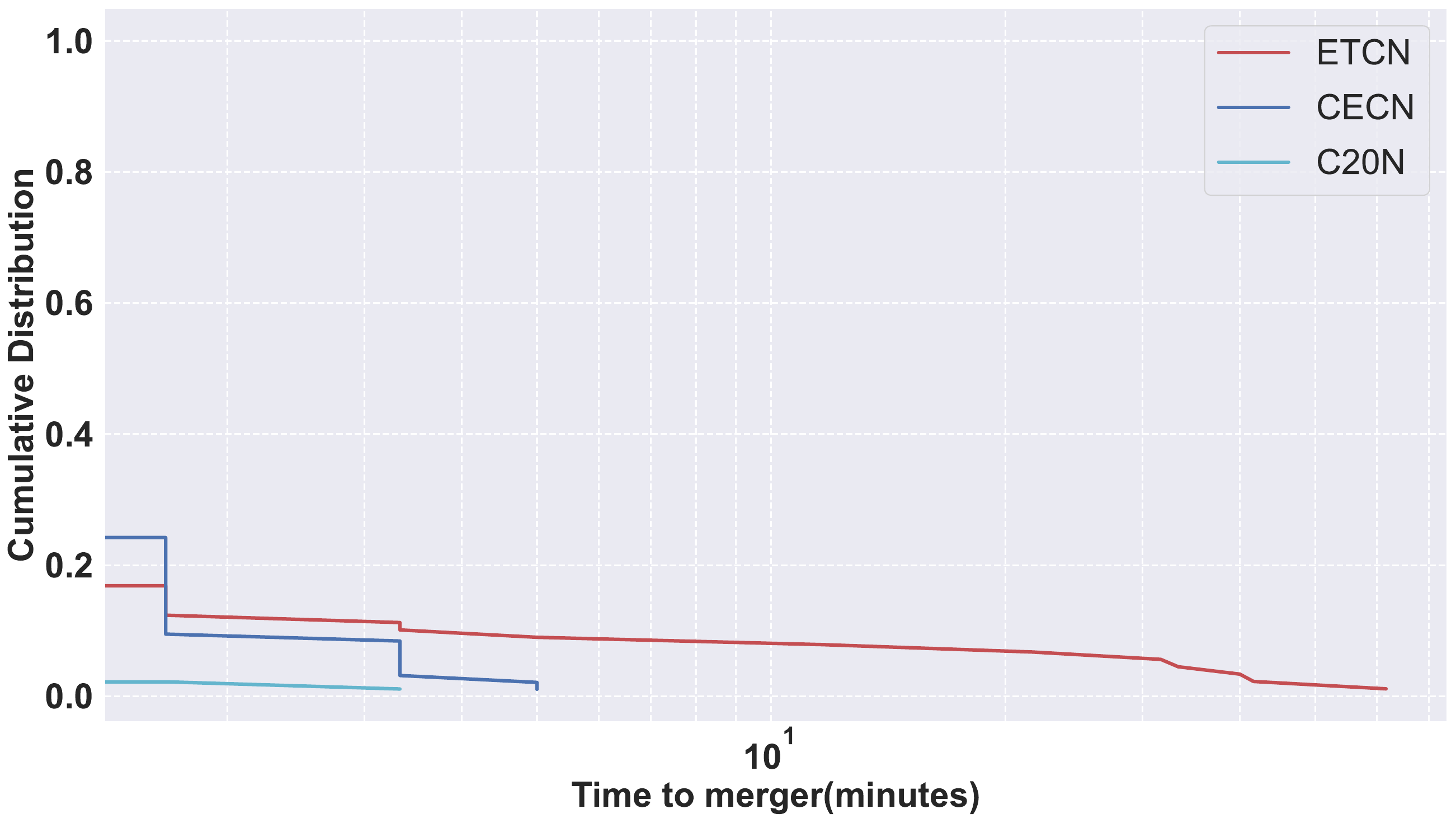}} 
\caption{The cumulative distribution of time to merger for \ac{BNS}s at 1600 Mpc.}\label{1600EW}
\end{figure}

\begin{figure}
\subfigure[ETCN network] { 
\includegraphics[width=1\columnwidth]{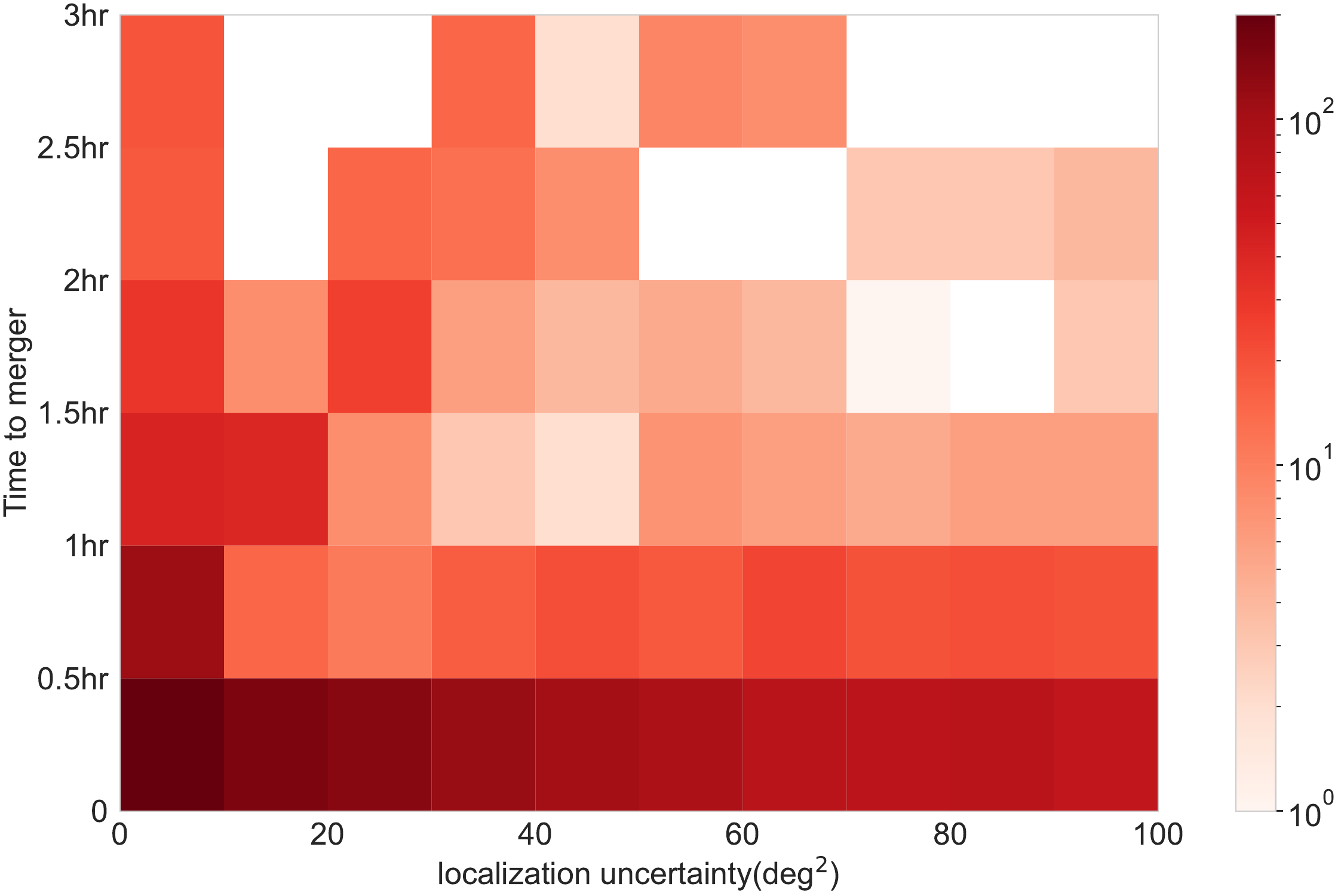}} 
\subfigure[CECN network] { 
\includegraphics[width=1\columnwidth]{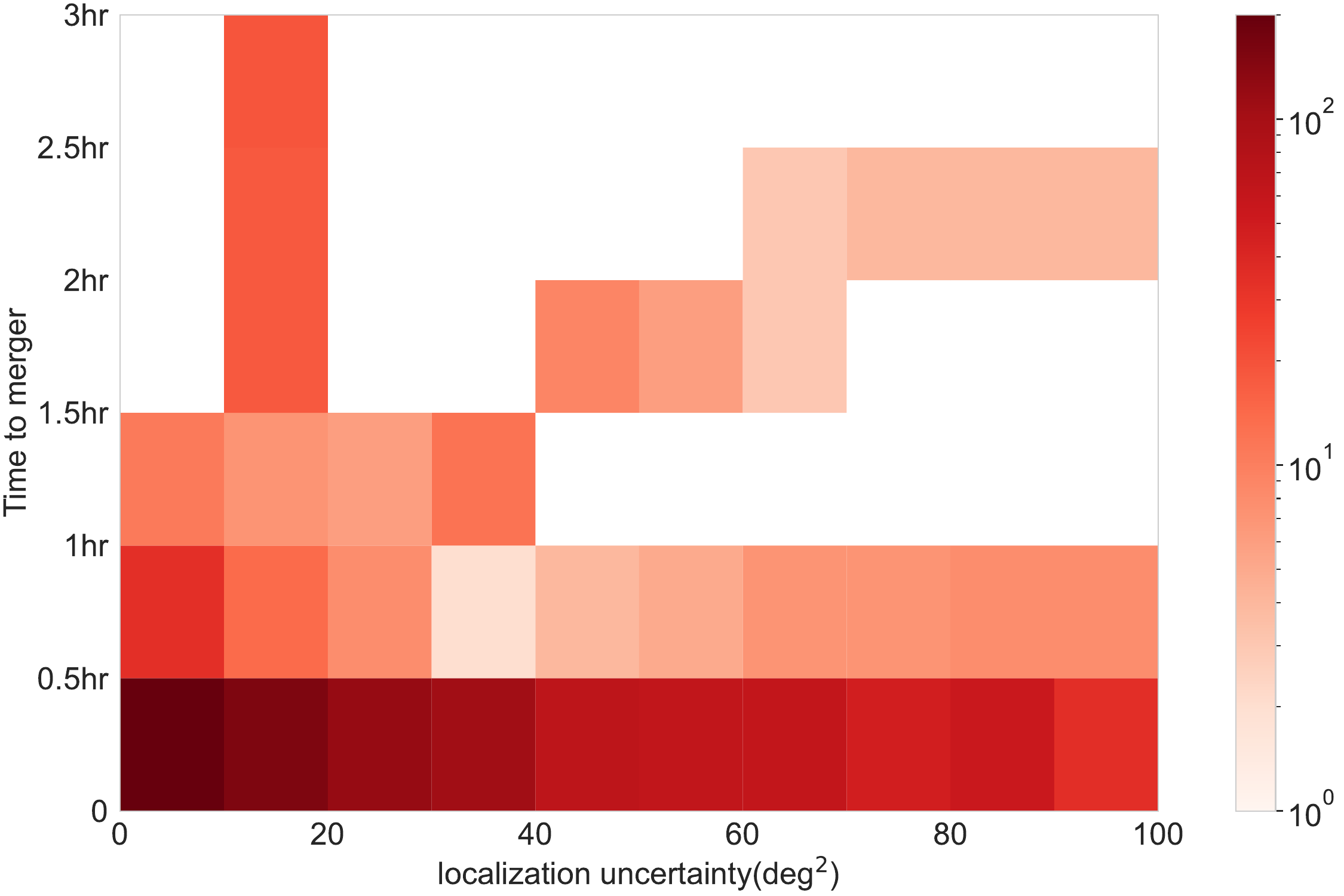}} 
\subfigure[C20N network] { 
\includegraphics[width=1\columnwidth]{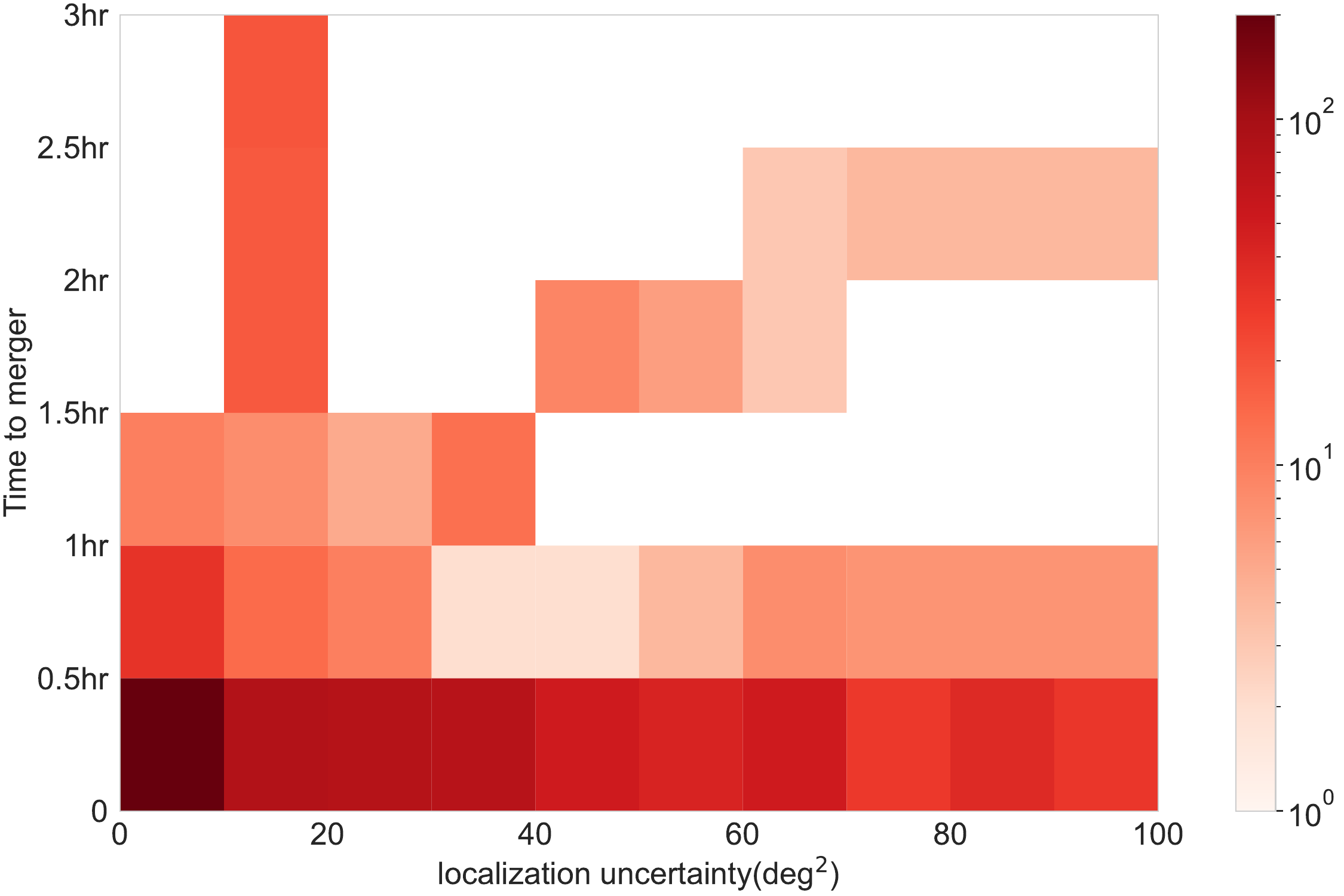}} 
\caption{Two-dimensional histograms showing the distributions of the time to merger for \ac{BNS}s in Simulation I. The horizontal axis of every subplot is the size of the 90\% credible region, and the vertical axis represents the time to merger. The color represents the number of sources that achieve the early warning criteria with the localization requirement being $\leq$ 100 deg$^2$.}
\label{EW_3G_appendix}
\end{figure}
\label{lastpage}
\end{document}